\documentclass[reprint,superscriptaddress,amsmath,amssymb, aps, prl, floatfix]{revtex4-2}

\usepackage{graphicx} 
\usepackage{amsmath,amsthm, amssymb}
\usepackage{bbm}
\usepackage{physics}
\usepackage{xcolor}
\usepackage{time}
\usepackage[pdftex,colorlinks=true]{hyperref}
\usepackage[colorinlistoftodos]{todonotes}
\usepackage{amsfonts}
\usepackage{mathtools}
\usepackage{bm}
\usepackage{comment}
\usepackage{ragged2e}
\usepackage{color,soul}
\usepackage{todonotes}
\newcounter{todocounter}

\def \be{\begin{equation}}
\def \ee{\end{equation}}
\def \bea{\begin{eqnarray}}
\def \eea{\end{eqnarray}}

\begin{document}

\title{Attention-Based Foundation Model for Quantum States} 

\author{Timothy Zaklama}
\email{tzaklama@mit.edu}
\affiliation{Department of Physics, Massachusetts Institute of Technology, Cambridge, Massachusetts 02139, USA}

\author{Daniele Guerci}
\affiliation{Department of Physics, Massachusetts Institute of Technology, Cambridge, Massachusetts 02139, USA}

\author{Liang Fu}
\affiliation{Department of Physics, Massachusetts Institute of Technology, Cambridge, Massachusetts 02139, USA}

\begin{abstract}
We present an attention-based foundation model architecture for learning and predicting quantum states across Hamiltonian parameters, system sizes, and physical systems. Using only basis configurations and physical parameters as inputs, our trained neural network is able to produce highly accurate ground state wavefunctions.  
For example, we build the phase diagram for the 2D square-lattice $t$–$V$ model with $N$ particles, from only 18 training points in parameter space $(V/t,N)$. 
Thus, our architecture provides a basis for building a universal foundation model for quantum matter.
\end{abstract}

\maketitle

\section{Introduction}
\label{sec:intro}


Foundation models such as large language models have reshaped AI by learning general-purpose representations from broad, heterogeneous data.
In the physical sciences, analogous ambitions have driven foundation-style efforts in chemistry and materials science, where models are trained across diverse systems to predict properties at reduced computational cost~\cite{Bommasani2021,Vaswani2017,Brown2020,Radford2021,Ramesh2021,Jumper2021,Evans2022, Rende2025}. 
Most of these efforts rely on density-functional-theory (DFT) pipelines for electronic structure.
DFT centers on the ground-state density and thus omits details accessible only through the ground-state wavefunction, including the full complex structure and higher-order correlations. 
A foundation model that learns ground-state wavefunctions directly would enable a more unified and expressive description of matter across systems, geometries, and phases.

Neural-network quantum states (NQS) provide compact and expressive parameterizations of many-body wavefunctions and have achieved impressive accuracy in lattice models~\cite{CarleoTroyer2017,Carrasquilla_2017,luo2019backflow,chen2025neuralnetworkaugmentedpfaffianwavefunctions}, quantum chemistry~\cite{Pfau2020,foster2025ab}, and continuum electron systems~\cite{Smith2024, geier2025self,teng2025solving,nazaryan2025finding,gaggioli2025electronic}. Progress has come from architectural innovations—restricted Boltzmann machines~\cite{CarleoTroyer2017}, convolutional networks~\cite{Carrasquilla_2017,Liang2018solving}, autoregressive flows~\cite{Sharir2020}, and attention-based models ~\cite{Pfau2020,geier2025self,teng2025solving,nazaryan2025finding,gaggioli2025electronic}—and from robust optimization methods such as stochastic reconfiguration for variational Monte Carlo (VMC)~\cite{Sorella2017}. 
Notably, attention-based architectures have emerged into an unifying NQS framework capable of representing diverse quantum phases and achieving state-of-the-art variational energies  ~\cite{TQS2023,zhu2023hubbard,Viteritti2023Transformer,Psiformer2022,Viteritti_2025}. 
However, most studies remain system specific, and require computationally demanding calculations for each Hamiltonian.

Our objective is to learn a mapping from Hamiltonian parameter space to many-body wavefunction space: $\boldsymbol\lambda\mapsto\ket{\Psi_\theta(\boldsymbol\lambda)}$, predicting the full wavefunction, such as that of the ground state. Once this map is learned, energies and equal-time observables follow directly, and the system is effectively solved anywhere on the considered parameter neighborhood. 


To achieve this goal, we introduce \emph{Q-stage}, an attention-based neural network architecture for \emph{Q}uantum \emph{sta}te \emph{ge}neration. We choose to work in the second quantized basis where the network's input tokens are Fock space configurations and Hamiltonian parameters $\boldsymbol\lambda$. 
Importantly, tokens are embedded through positional encodings, which provide a efficient feature space representation of the system, and attention operations naturally captures quantum correlations. 
Translational symmetry is not imposed by construction; rather, spatial correlations are learned directly from sampling. This makes the architecture naturally applicable to models with spatial inhomogeneities, for instance onsite disorder.
We train the network on a small number of exact-diagonalization (ED) ground states to learn a smooth map $\boldsymbol\lambda\mapsto\ket{\Psi_\theta(\boldsymbol\lambda)}$, for different $N$. 
On the 1D and 2D $t$–$V$ model up to $16$ sites, the model attains more than $99.9\%$ overlap on training points and more than $99.5\%$ on unseen parameters, and matches ED energies within $1.5\%$. A handful of labeled Hamiltonians suffices to reconstruct the full $(V/t,N)$ phase diagram ($N$ is number of particles), interpolating seamlessly through phase transitions and Hilbert space sectors. 

We show that physics insight into training data selection is crucial for generalization, and the success of the foundation model relies on a smooth mapping from the parameter space to the wavefunction space. 
Our method is highly flexible and can be adapted to other setting beyond learning just quantum ground states. Limitations do exist: the wavefunction is discontinuous across first order phase transitions and excited-state are more difficult to learn than ground states. On the whole, our results demonstrate that a foundation model can be achieved for learning quantum states across different Hamiltonian parameters, models, and physical systems.

\section{Theoretical Framework}
\label{sec:TheoM}

Let $\{\hat H_{\boldsymbol\lambda}\}$ denote a family of quantum many-body Hamiltonians indexed by couplings $\boldsymbol\lambda \in \Lambda$ (e.g., $\boldsymbol\lambda=(V/t,N)$ for the $t$--$V$ model at particle number $N$). 
For each $\boldsymbol\lambda$, let $\mathcal H_{\boldsymbol\lambda}$ be the corresponding Hilbert space with orthonormal Fock basis $\{\ket{\mathbf n}\}$, where $\mathbf n=(n_1,\dots,n_{L\times L})$, and $n_i$ denotes occupation of the $i^{th}$ mode, for example, the $i^{th}$ site on a lattice of $N_s=L\times L$ sites.
Our architecture aims to find single parameter-sharing map $\psi_\theta(\mathbf n ;\boldsymbol\lambda)$, such that 
\begin{equation}
\ket{\Psi_\theta(\boldsymbol\lambda)} \;=\; \sum_{\mathbf n} \psi_\theta(\mathbf n ;\boldsymbol\lambda)\,\ket{\mathbf n},
\label{eq:ansatz}
\end{equation}
with shared parameters $\theta$ across all $\boldsymbol\lambda$. Achieving this map would realize a foundation model for quantum states.

We realize this map by minimizing a fidelity loss aggregated over an ensemble of Hamiltonians, from a finite training set $\mathcal S\subset\Lambda$. Let $\ket{\Psi_{\mathrm{ED}}(\boldsymbol\lambda)}$ denote the ED ground state for $\hat H_{\boldsymbol\lambda}$.
\begin{align}
\mathcal L(\theta) 
&=  1 - \sum_{\boldsymbol\lambda\in \mathcal S} w_{\boldsymbol\lambda}\,
\mathcal O_v(\boldsymbol\lambda),
\qquad
\sum_{\boldsymbol\lambda\in\mathcal S} w_{\boldsymbol\lambda}=1,
\label{eq:loss-disc}
\end{align}
for 
\begin{equation}
    \mathcal O_v(\boldsymbol\lambda) 
    = \frac{\bigl|\langle{\Psi_\theta(\boldsymbol\lambda)}|{\Psi_{\mathrm{ED}}(\boldsymbol\lambda)}\rangle\bigr|^2}{\sqrt{\|\Psi_\theta(\boldsymbol\lambda)\|^2\,\|\Psi_{\mathrm{ED}}(\boldsymbol\lambda)\|^2}}.
    \label{eq:overlap}
\end{equation}

By construction, $0\le \mathcal L(\theta)\le 1$, and $\mathcal L(\theta)=0$ if 
$\ket{\Psi_\theta(\boldsymbol\lambda)}$ matches $\ket{\Psi_{\mathrm{ED}}(\boldsymbol\lambda)}$ up to a $\boldsymbol\lambda$-dependent global phase for all $\boldsymbol\lambda\in\mathcal S$.
The ensemble weights or $\{w_{\boldsymbol\lambda}\}$ may be adapted to the application: uniform over a training set, or non-uniform. Because $\theta$ is shared and each state knows about $\boldsymbol\lambda$, the same model can be trained on arbitrarily many Hamiltonians in a single run by increasing the size of $\mathcal S$.

\section{Model Architecture}
\label{sec:ModArch}

\begin{figure}
    \centering
    \includegraphics[width=1\linewidth]{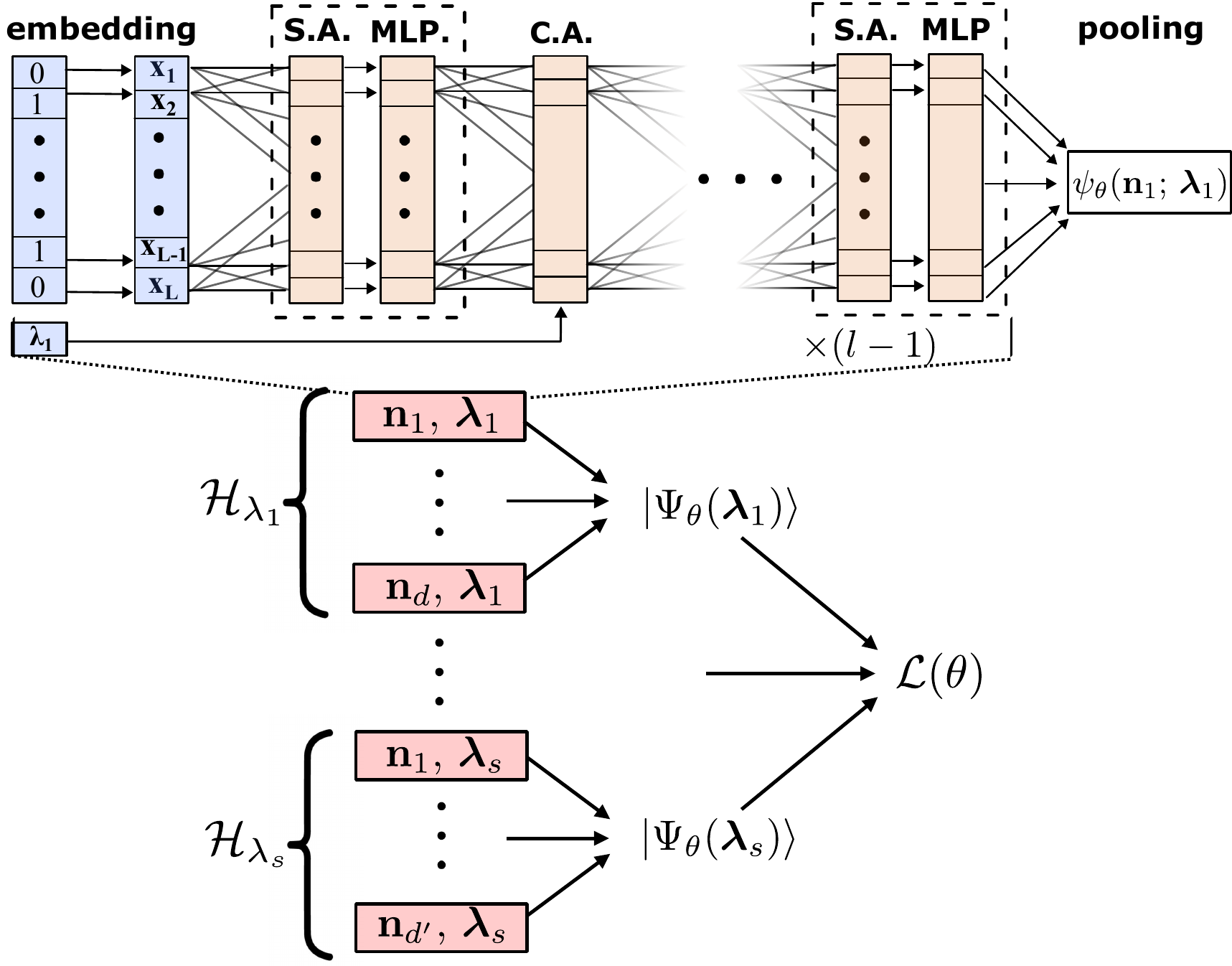}
    \caption{\textbf{Architecture of the neural network wavefunction ansatz.} {Our model first embeds an input Fock state, then passes the token through a self attention block of finite depth, followed by a single cross attention block, whereby the token is finally pooled to the complex basis state output. The complex amplitude for each basis state is then aggregated into the wavefunction for a particular set of Hamiltonian parameters $\boldsymbol \lambda$, which are all compiled into loss function. Finally, gradient descent is performed using adamw optimizer, and the model finishes training once a sufficient average overlap is reached.}}
    \label{fig:arch}
\end{figure}

Our architecture is illustrated in Fig. \ref{fig:arch}. We denote each Fock basis state by the
binary vector $\mathbf n=(n_1,\dots,n_L)\in\{0,1\}^L$ and the Hamiltonian parameters by
$\boldsymbol\lambda=(V/t,N)\in\mathbb R\times \mathbb N$. The network produces a complex amplitude for each basis state
\begin{equation}\label{parametrization}
    \psi_\theta(\mathbf n;\boldsymbol\lambda)\;=\;
    c_\theta^{\mathrm{Re}}(\mathbf n;\boldsymbol\lambda) \;+\;
    i\,c_\theta^{\mathrm{Im}}(\mathbf n;\boldsymbol\lambda)\;,\qquad c_\theta\in\mathbb R,
\end{equation}
and the many-body wave function is given in Eq.~\eqref{eq:ansatz} for a given $\boldsymbol{\lambda}$.
In contrast to other transformer architectures and foundation models that uses patching and sequence processing to parse inputs \cite{Rende2025,Viteritti2023Transformer,Viteritti_2025, ZhangDiVentra2023}, we choose to embed using positional encoding of the sites in our basis states.
Our architecture has the following simple structure:

\begin{enumerate}
    \item Each input token representing a Fock basis state, $|\mathbf{n}\rangle$, is \textit{embedded} into a high dimension space as a vector $\mathbf{X} = (\mathbf{x}_1,...,\mathbf{x}_{L\times L})$ with $\mathbf{x}_i$ encoding both the occupation and position of site $i$.
    \item Each embedded token, $\mathbf{X}$, is then passed through a \textit{self attention block} to mix the embedded per site elements $\mathbf{x}_1,...,\mathbf{x}_L$.
    \item Each token is further passed through a \textit{cross attention block} to couple to the Hamiltonian parameters $\boldsymbol{\lambda}$.
    \item Finally, each token is projected back to the wavefunction output dimension through \textit{pooling}.
\end{enumerate}

\emph{Embeddings.}
Each basis state is assigned a token $\mathbf X$, which is built from embedding each site to a higher dimensional space (feature space of which the dimension is a hyper parameter) as follows: 
Each site occupation number $n_i$ is mapped to a $f$-dimensional ($f$ for feature) vector by a learnable embedding map $e$, and a learned positional vector (gaussian initialized weights),
\begin{equation}
    \mathbf x_i \;=\; e(n_i)\;+\;\mathbf p_i,\qquad e:\{0,1\}\to\mathbb R^{f},\;\; \mathbf p_i\in\mathbb R^{f}, \label{embedding}
\end{equation}
where the feature function $e$ is site-independent, while the positional vector $\mathbf p_i$ depends explicitly on the site index. Importantly, $\mathbf{x}_i$ encodes both the occupation and the position of each site on the lattice.

Collecting vectors $\mathbf{x}_1,...,\mathbf{x}_L$ yields the token for each basis state $\mathbf X=(\mathbf x_1,\dots,\mathbf x_L)\in\mathbb R^{L\times f}$. Thus the full token space is of dimension $\mathbb R^{d \times L\times f}$, where $d$ is the total number of basis states, and each basis state is processed as an independent single state stream.

\emph{Self-attention Block.}
A stack of $D$ (depth) self-attention blocks updates $\mathbf X$; each block undergoes multi-head attention (MHA) followed by a Multi-Layered Perceptron (MLP), both with residual connections \cite{He2016DeepResNet}:
\begin{equation}
    \begin{aligned}
        \tilde{\mathbf X} &= \mathbf X_{in} \;+\; \mathrm{MHA}\!\bigl(\mathrm{LN}(\mathbf X_{in})\bigr),\\
        \tilde{\mathbf X} &\;+\; \mathrm{MLP}\!\bigl(\mathrm{LN}(\tilde{\mathbf X})\bigr) \rightarrow \mathbf X_{out},
    \end{aligned}
\end{equation}
where LN is layer normalization, which retains permutation equivariance, stabalizes learning, and improves convergence without sacrificing significant memory \cite{Ba2016LayerNorm}.

With $H$ heads and per-head width $w_h=f/H$, the explicit equation for multi-head self-attention is standard:
\begin{multline}
    \mathrm{MHA}(\mathbf X)\;=\;\Bigl[\big\|_{h=1}^{H}
    \underbrace{\mathrm{softmax}\!\Bigl(\tfrac{\mathbf Q_h \mathbf K_h^\top}{\sqrt{w_h}}\Bigr)\mathbf V_h}_{\text{head }h}\Bigr]\,W^{O}, \\
    \mathbf Q_h=\mathbf X W_h^{Q},\;\mathbf K_h=\mathbf X W_h^{K},\;\mathbf V_h=\mathbf X W_h^{V},
\end{multline}
and the position-wise MLP acts identically and \textbf{independently} at each site:
$\mathrm{MLP}(\mathbf X)=W_2\,\phi(W_1\mathbf X+\mathbf{b}_1)+\mathbf{b}_2$, for some non-linear function $\phi$.

\emph{Cross-attention Block.}
To co-train across $(V/t, N)$ we form a single parameter token
\[
\mathbf p_{\lambda}=W_\lambda\,\boldsymbol\lambda\in\mathbb R^{f},\qquad
\mathbf P_{\lambda}=[\mathbf p_{\lambda}]\in\mathbb R^{1\times f},
\]
and inject it after the self-attention stack via cross-attention where the basis state token, $\mathbf{X}$ are
queries and the parameter token supplies keys/values:
\begin{multline}
    \mathbf X_{in} \;+\;
    \Bigl[\mathrm{softmax}\!\Bigl(\tfrac{\mathbf Q\,\mathbf K_\lambda^{\!\top}}{\sqrt{E}}\Bigr)\mathbf V_\lambda\Bigl]W^O \rightarrow \mathbf X_{out},\\
    \mathbf Q=\mathbf X W^{Q},\;\;
    \mathbf K_\lambda=\mathbf P_\lambda W^{K},\;\;
    \mathbf V_\lambda=\mathbf P_\lambda W^{V}.
\end{multline}
With a single parameter token, cross-attention broadcasts a learned projection of $(V/t,N)$ to every site, independent of the basis state. Every site, and therefore every state is conditioned on uniformly injected $\lambda=(V/t,N)$, and all subsequent self-attention layers use it to modulate amplitudes. Crucially, cross-attention is only passed through once, as the parameter label need only to tag each state once. Multiple injections of cross-attention would introduce redundancy and occupy valuable memory. We elected to place cross attention after the first self attention layer to maximize network conditioning while also to streamline learning (see supplementary materials (SM)~\cite{supplementary} for deeper explanation).

\emph{Pooling and readout.}
Global permutation-equivariant information is obtained by MLP pooling over sites, with Hamiltonian parameters $(V/t,N)$ concatenated to each basis state:  
\begin{equation}
        \mathbf h=\text{MLP}(\{\mathbf{X}, V/t, N \}) \in \mathbb R^{f},\qquad \widehat{\mathbf{h}}=\mathrm{LN}(\mathbf h).
\end{equation}

The pooling is then followed by a linear readout to $(c^{\mathrm{Re}},c^{\mathrm{Im}})$:
\begin{equation}
    (c^{\mathrm{Re}},c^{\mathrm{Im}})\;=\;\widehat{\mathbf h}\,W_{\mathrm{out}}+ \mathbf b_{\mathrm{out}}\in\mathbb R^{2}, \label{eq:readout}
\end{equation}
where, for simplicity, we drop the explicit dependencies of the coefficients $(c^{\mathrm{Re}},c^{\mathrm{Im}})$ in Eq.~\eqref{parametrization}.

Pooling reduces the latent model dimension to the output dimension usually via averaging and normalizing, projecting the token to the output dimension. Here, we pool by way of MLP, which is far more efficient at compressing the information, revealing another key innovation of this work. This readout (Eq. \ref{eq:readout}) realizes $\psi_\theta(\mathbf n;\boldsymbol\lambda)=c^{\mathrm{Re}}_\theta(\mathbf n;\boldsymbol\lambda)+i\,c^{\mathrm{Im}}_\theta(\mathbf n;\boldsymbol\lambda)$.   

It is important to note that since the network operates on each basis state token, $\mathbf{X}$, independently, the model does not depend on the ordering of the basis states. Critically, the model is able to learn the positions of individual sites through positional embeddings, $\mathbf p_i$ in Eq. \eqref{embedding}. In contrast to previous transformer NQS \cite{ZhangDiVentra2023, Viteritti2023Transformer, Viteritti_2025, Rende2025}, no patching or recurrent components are used to encode the ordering of the elements in the basis states. As transformers are designed to take in a discrete input, the occupational basis used here is especially appropriate and our transformer architecture is thus both maximally and naturally suitable to represent ground state wavefunctions on lattices.


\begin{figure}
    \centering
    \includegraphics[width=\linewidth]{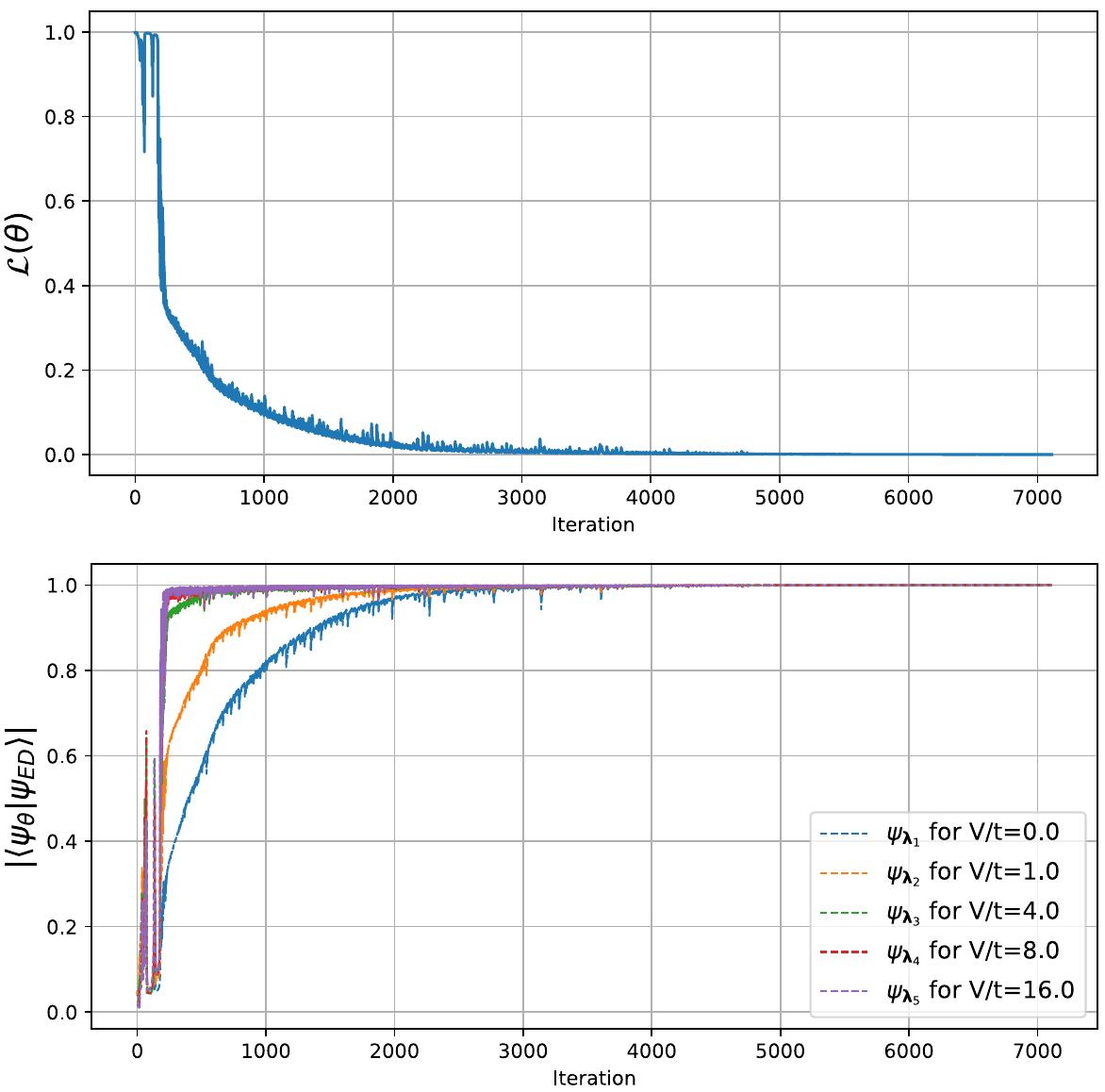}
    \caption{\textbf{Loss and cumulative overlap during training for 2D square lattice ($N_s=16, N=8$).} {Despite training on a wide range of Hamiltonian parameters the loss function asymptotically approaches 0 (total fidelity approaches 100$\%$). The non-interacting ground state is the hardest to learn; nevertheless it is eventually fully learned and asymptotically approaches a fidelity of 1.}}
    \label{fig:loss}
\end{figure}

\begin{figure}
    \centering
    \includegraphics[width=\linewidth]{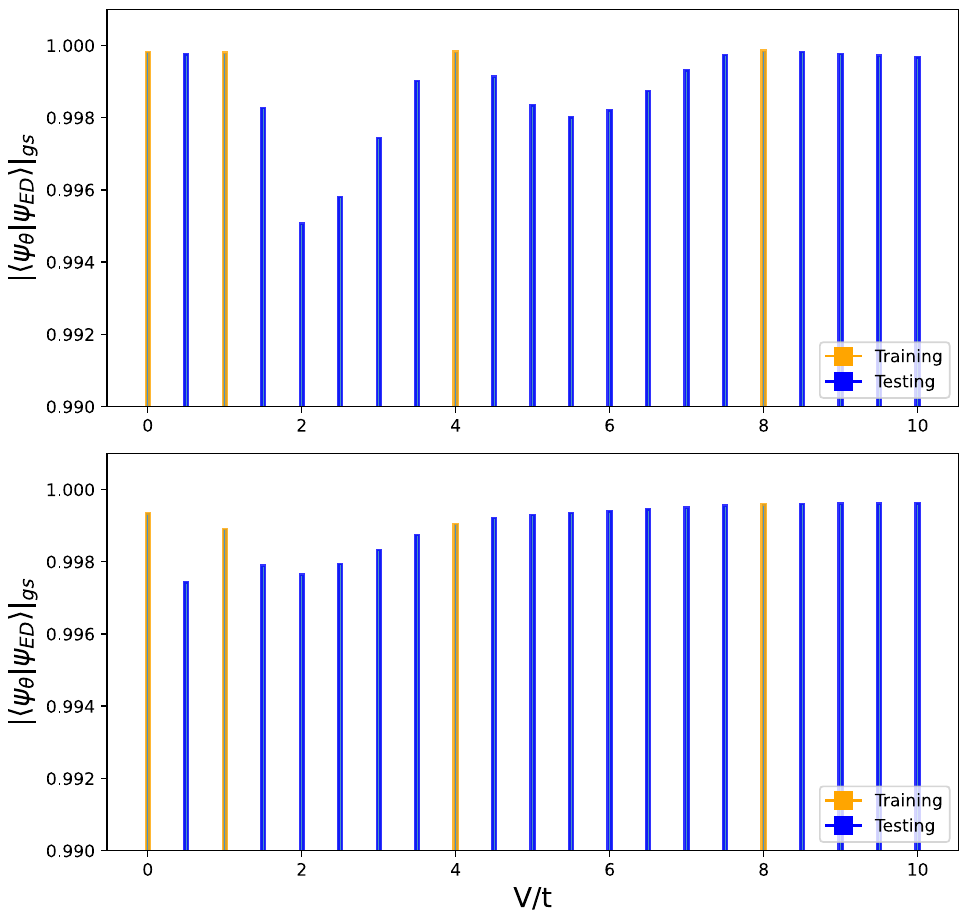}
    \caption{\textbf{Overlap generalization over parameter space ($V/t$) for $N_s=16, N=8$}, in 1D (top) and 2D (bottom). {Average overlap is trained until 99.99$\%$ fidelity is reached. Generalization exceeds 99.5$\%$ fidelity for all cases tested.}}
    \label{fig:overlap-bars}
\end{figure}

\section{Physical Model}
\label{sec:PhysM}

We consider spinless fermions on a lattice with nearest-neighbor hopping and repulsion: the $t$–$V$ model. On lattice $\text{L}$  with periodic boundary conditions,
\begin{equation}\label{hamiltonian}
    \hat H \;=\; -\,t \sum_{\langle r,r'\rangle}\! f^{\dagger}_{r} f_{r'} \;+\; V \sum_{\langle r,r'\rangle}\! \rho_{r}\,\rho_{r'}, 
    \quad \rho_r=f^{\dagger}_r f_r,
\end{equation}
where $t$ is the hopping amplitude and $V$ the nearest-neighbor interaction. 
We write the filling as $\nu = N/N_s$ with $N_s$ number of sites. 
This minimal model captures the essence of certain moir\'e semiconductors, and exhibits a rich phase diagram phase including charge order and superconductivity ~\cite{Slagle2020charge,crepel2021new,guerci2025ferromagneticsuperconductivityexcitoniccooper,YangZhang2021}.

In 1D, the $t$–$V$ model maps to the XXZ spin chain and is exactly solvable by Bethe ansatz and bosonization. 
At weak coupling ($V/t\!\to\!0$) the ground state is a metallic Luttinger liquid with power-law density correlations; at strong coupling the ground state develops charge-density-wave (CDW) order and becomes insulating. 
At half filling, the transition between the Luttinger liquid and the CDW phase is of the Kosterlitz–Thouless type and occurs at $V/(2t)=1$~\cite{Haldane1982,Mila_1993}.

\begin{figure*}
    \centering
    \includegraphics[width=\linewidth]{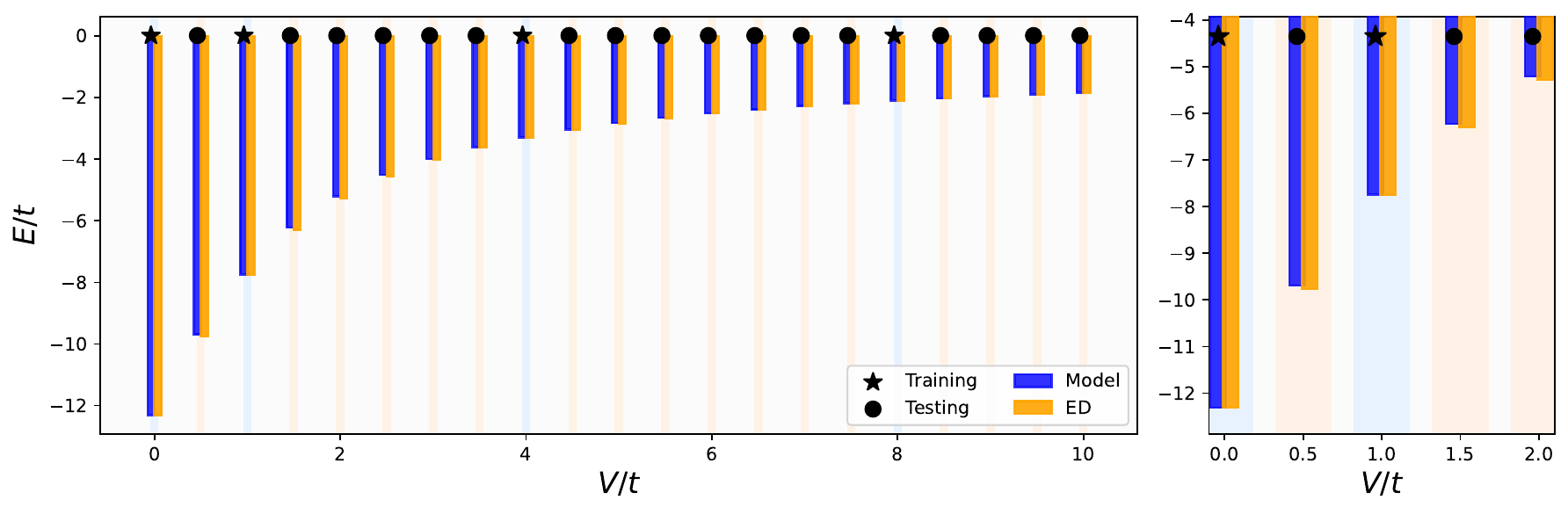}
    \caption{\textbf{Energy generalization over parameter space ($V/t$) for $N_s=16, N=8$ in 2D.} {Energy agrees to 2 digits and has a percent error of under 1.5$\%$ for all cases. There is no significant difference between training and test cases. 
    The right panel highlights the small-$V/t$ regime, which harder to train but still reproduced with high accuracy.}}
    \label{fig:energy-bars}
\end{figure*}
 
On the 2D square lattice, the phenomenology resembles the 1D case, but key differences arise: at half-filling, Fermi-surface nesting in the non-interacting system produces a weak-coupling instability toward a checkerboard CDW insulator for arbitrarily small repulsion $V/t$~\cite{Gubernatis1985,Stokes2020,Czart2008}. 
Doping away from half-filling frustrates charge order and promotes superconductivity~\cite{He2023superconductivity}.
Determining the balance between these two competing tendencies remains a nontrivial numerical challenge and has been the focus of several studies~\cite{Stokes2020, He2023superconductivity, Czart2008, deWoul2010, Song2014, Sikora2015, Wang_2014}.

We use periodic boundaries and even $N_s$ so that half filling is supported.
Our benchmarks span both 1D rings and 2D squares; the principal results reported below focus on the 2D model at fixed {$N_s$} and varying $(V/t,N)$, matching the training/evaluation grids used for the neural wavefunction experiments.

To obtain a unique, symmetry–consistent ground state at each parameter point—required for the network to learn the smooth $\boldsymbol \lambda \rightarrow \ket{\Psi_\theta (\boldsymbol \lambda)}$ map—we must resolve the exact degeneracies of the square–lattice \(t\!-\!V\) model with a small symmetry–selective perturbation. 
We then produce ED wavefunctions from the augmented Hamiltonian
\begin{equation}
    \hat H_{\mathrm{ED}}
    \;=\;
    \hat H
    \;+\;
    \hat H_{\mathrm{sub}},
\end{equation}
where \(\hat H_{\mathrm{sub}}\) breaks the sublattice degeneracy. 
Residual degeneracies are resolved utilizing symmetry eigenvalues (see the SM \cite{supplementary} for details). Explicitly, 
\begin{equation}
    \hat H_{\mathrm{sub}}
    \;=\;
    \delta \sum_{r} \eta_r\,\rho_r,
    \qquad
    \eta_r = (-1)^{x_r+y_r},
\end{equation}
a checkerboard potential that lifts the sublattice degeneracy (we set \(\delta = 0.01\)), with \(x_r\) and \(y_r\) the integer lattice coordinates of site \(r\).

Resolving degeneracies is necessary for the network to learn a smooth mapping $\boldsymbol \lambda \rightarrow \ket{\Psi_\theta (\boldsymbol \lambda)}$. Without a smooth map, the universal approximation theorem fails, and the network cannot generalize. Our approach thus selects the appropriately unique ground states over the entire $(V/t,N) \rightarrow \ket{\Psi_\theta(V/t,N)}$ parameter {space} that our model can learn continuously. As such, a foundation model will not be able to interpolate across first order phase transitions, and these points, in general, need to be handled with more care. This expressivity constraint is a general principle of foundation models for quantum states and applies to any Hamiltonian or classes of Hamiltonians.

\section{Results}
\label{sec:res}

Our results were produced using LeCun Initialization, adamw optimzer with weight decay of $3\times 10^{-3}$, linear learning rate decay schedule of $1\times 10^{-3}$ to $1\times 10^{-4}$ after 95$\%$ overlap is reached, network hyperparameters of 2 layers, feature dimension of 80, MLP width of 80 with 1 hidden layer, and 8 attention heads. No hyperparameter tuning, gradient clipping, explicit regularization, or additional complexity was introduced.

We train a single wavefunction $\ket{\Psi_\theta(\boldsymbol\lambda)}$ on a small set of Hamiltonians using no Monte Carlo sampling and optimizing the wavefunction directly against that of ED. 
We report (i) the overlap (Eq. \ref{eq:overlap}),
(ii) the ground-state energy $E_{\theta}$ and relative error $\Delta E/E
=|E_\theta-E_{\rm ED}|/|E_{\rm ED}|$, and (iii) ground state versus excited state learning.  
For the 2D $4{\times}4$ square case, we first train on four distinct $V/t$ parameter points at fixed $N=8$ and smoothly generalize across all $V/t$. 
We then train on 18 $(V/t,N)$ points and generalize across the entire $(V/t,N)$ plane, generating more than $150$ quantum ground states—including regions lying outside the fixed-particle subspace—without any change to the model. 
Our phase diagram reinforces that of previous studies~\cite{Stokes2020}, and is the first NQS model to generalize over the $(V/t,N)$ parameter space with a finite training set.

Figure~\ref{fig:loss} shows a smooth, monotonic loss decay with rapid convergence to near-perfect fidelity: the average training overlap reaches $>\!99.99\%$. 
The non-interacting ground state ($V/t{=}0$) is the hardest wavefunction to learn, exhibiting a slower convergence branch in Fig.~\ref{fig:loss} that coincides with the onset of a nontrivial Fermi-surface structure. 
This “easy-to-hard” trajectory illustrates an intuitive learning curriculum: coarse, global features are learned first, allowing the network to focus on fine-grained amplitude/phase structure in later stages. 
As a result, we note the importance of allowing the network to reach a high convergence ($>\!99.99\%$), as generalization grows disproportionately as the average overlap plateus to 1.

Despite training on only five Hamiltonians, the model achieves striking test generalization across the full $V/t$ axis in both one and two dimensions. 
Figure~\ref{fig:overlap-bars} reports $\mathcal{O}_v\!>\!99.5\%$ on every held-out $V/t$ value. 
In 2D the learning is modestly slower than in 1D (consistent with the more complicated phase manifold), yet the final generalization level is actually slightly better (indicating the importance of training point selection). 
These results indicate that the model has learned a highly data-efficient map $\,\boldsymbol\lambda\!\mapsto\!\ket{\Psi_\theta(\boldsymbol\lambda)}\,$ from very few parameter inputs.

Although the model never minimizes energy directly, it not only pinpoints the ground state energy, but also generalizes with high prediction accuracy. Figure~\ref{fig:energy-bars} shows in and out-of-sample ground-state energies within $1.5\%$ of the exact value across the tested grid. Thus, the ground state energy can be predicted for all Hamiltonians within a class from a minimal training subset. More broadly and powerfully, \emph{any} equal-time observable for \emph{any} set of Hamiltonian parameters can be straightforwardly calculated. The accuracy is guaranteed from learning the full complex wavefunction ($\ket{\Psi_\theta(\boldsymbol\lambda)}$), which means once the model is trained, every equal-time observable can be obtained in post-processing.
Thus, training on wavefunctions, rather than local energies, yields broad observable accuracy after a single supervised pass. 
Even further, the network could in principle be fine-tuned to produce even more accurate measures by simply importing the weights of the model, then further optimizing on whatever target-such as a correlation function or another observable to the loss function-if desired.

\begin{figure}
    \centering
    \includegraphics[width=\linewidth]{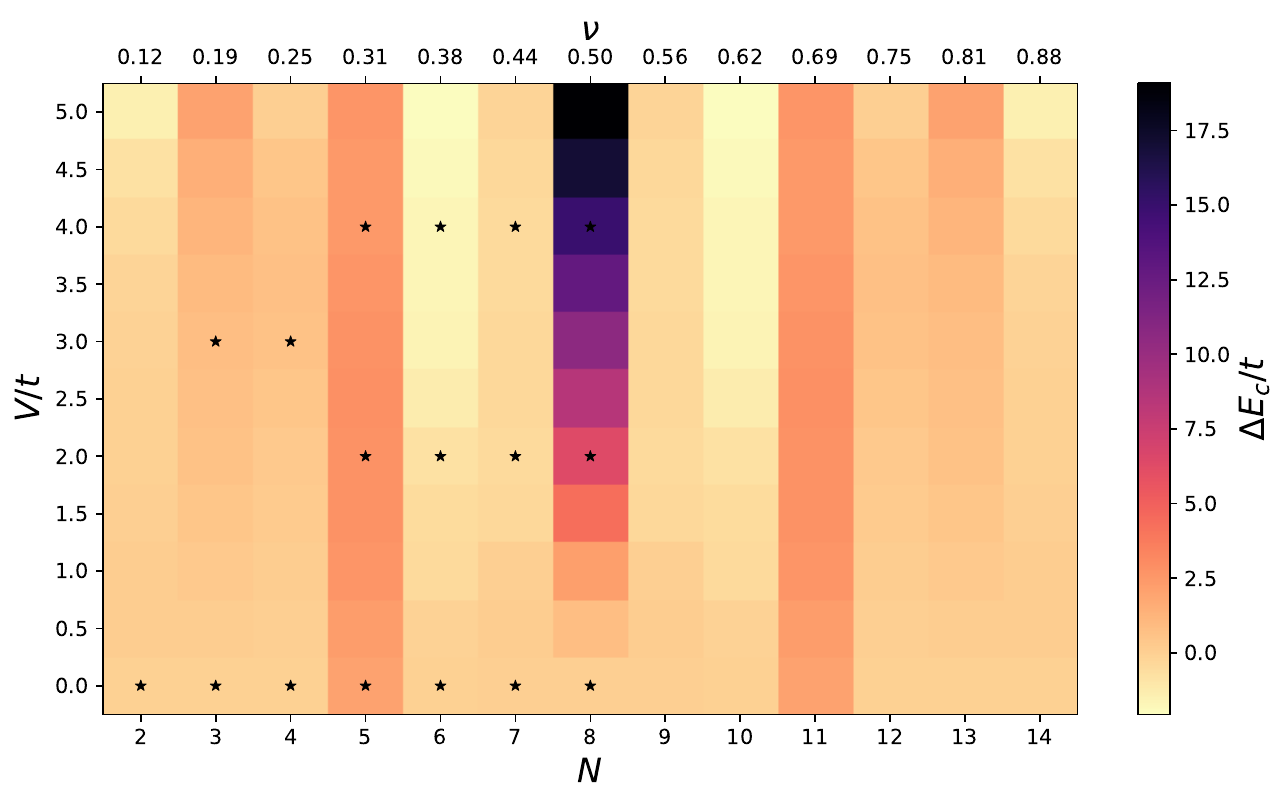}
    \caption{\textbf{Charge excitation energy heatmap of 2D $t$--$V$ model ({$N_s=16$})}. Black stars denote training points. The entire phase diagram is accurately predicted from just a handful of training points.}
    \label{fig:chargeGap}
\end{figure}

Directly applying the generalization power of our model to learn ground states, we plot the charge excitation energy for the entire parameter space of our model in Fig. \ref{fig:chargeGap}, revealing the phase structure. 
We define this quantity as:
\begin{equation}
    \Delta E_c(N)= E_{N+1} +  E_{N-1} - 2E_{N},
\end{equation}
where energy $E_N$ denotes the ground state energy for the system with $N$ particles. The energy $\Delta E_c(N)$ in Fig.~\ref{fig:chargeGap} is symmetric about half filling ($N=8$), reflecting the particle–hole symmetry of the Hamiltonian~\eqref{hamiltonian}.

At half filling, $\Delta E_c \ge 0$ and extrapolates to $\Delta E_c \!\approx\! 0$ at $V/t{=}0$, consistent with a compressible metallic state. As $V/t$ increases, $\Delta E_c$ grows monotonically, signaling the crossover into the insulating checkerboard CDW phase, and approaches the asymptotic value $4V$ at large $V$.

Remarkably, upon doping away from half filling, an effective attraction develops: the charge excitation energy becomes negative ($\Delta E_c < 0$) at the particle–hole conjugate configurations $N=7,9$ and $N=6,10$ relative to the half filled state.
This implies that the energy of a pair of electrons to the ground state, $E_{2e} = E_{N+2} - E_{N}$, is lower than twice the energy of a single electron, $E_{1e} = E_{N+1} - E_{N}$ (i.e., $E_{2e} < 2E_{1e}$), the difference defines the pair binding energy $E_{b}(N)=E_{2e}-2E_{1e}=E_{N+2}+E_{N}-2E_{N+1}=\Delta E_{c}(N+1)$, indicating a net attractive interaction when doping the CDW insulator on a bipartite lattice~\cite{Slagle2020charge,crepel2021new,guerci2025ferromagneticsuperconductivityexcitoniccooper}. 
This behavior emerges despite the purely repulsive nature of the Hamiltonian~\eqref{hamiltonian} and is a consequence of the effective pairing interaction generated by electron–hole fluctuations of the CDW~\cite{Slagle2020charge,crepel2021new,guerci2025ferromagneticsuperconductivityexcitoniccooper}.  

Our neural ansatz is thus not merely accurate at isolated parameter points, but predictive over the entire square-lattice \(N\!-\!V/t\) phase diagram. In this sense, our approach is unified, and goes beyond single-parameter NQS calculations of ground state observables (a full phase diagram built from charge ordered correlations is provided in the SM~\cite{supplementary}).


\begin{figure*}
    \centering
    \includegraphics[width=\linewidth]{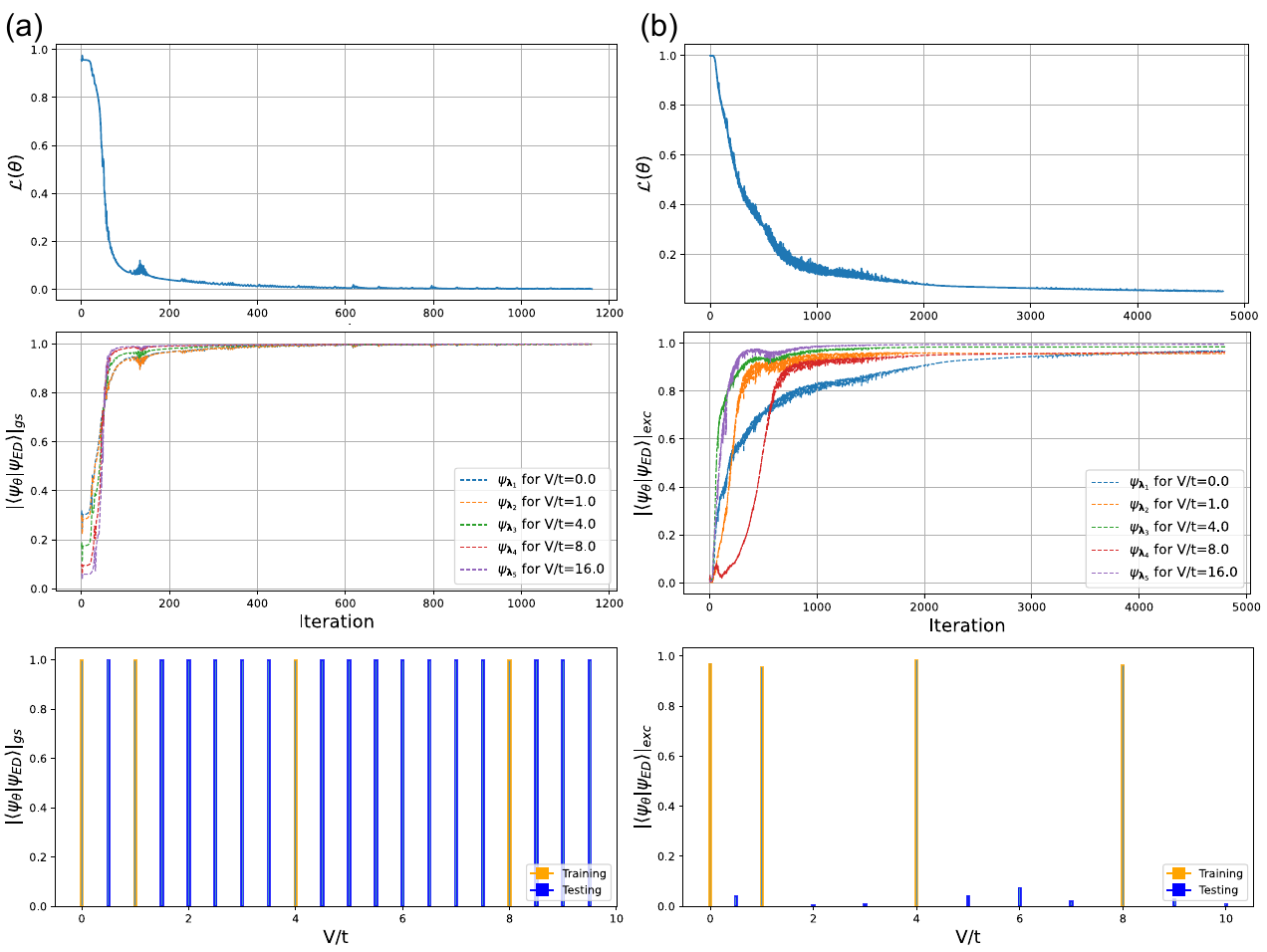}
    \caption{\textbf{Ground state (a) and mid-spectrum excited state (b) comparison for 1D {${N_s=12, N=6}$} system.} {The top figures are the loss curve, followed by the middle overlap plot of each wavefunction during training, with an generalization comparison at the bottom. The loss and overlap curves reveal the difficulty of learning and optimizing over excited states, while the overlap histogram demonstrates the failure of excited states to generalize versus their ground state counterparts.}}
    \label{fig:excited}
\end{figure*}

We also compare ground state to excited state learning by performing supervised optimization on a mid-spectrum state, and comparing the optimization and generalization to the corresponding ground state case. Figure~\ref{fig:excited} compares ground-state and mid-spectrum-excited-state learning for 1D 12 site chain at half filling with all the same network and optimization parameters. Despite training for over quadruple the number of iterations, the excited state final convergence is considerably higher than it's ground state counterpart (about $\mathcal{L}(\theta)^{converged}_{exc} = 0.1$ compared to $\mathcal{L}(\theta)^{converged}_{gs} = 0.001$). Further, the perfect generalization is lost in the excited state case, where the model completely fails to generalize at all. As expected, excited states generalize less robustly, as mid-spectrum states have extremely small level spacings, frequent level crossings, and higher sensitivity to boundary conditions across $(V/t,N)$. In this way, the network is learning a target state with rapidly changing properties over the parameter {neighborhood}, almost like trying to fit a moving target.
These results reveal an essential requirement for foundation model learning and generalization: the network must learn a smooth mapping between the target states and parameter space; if the mapping is sufficiently not smooth, the universal approximation theorem may not hold, and the network will have difficulty learning between wavefunctions and lose the ability to generalize. 
Thus, despite poor generalization for a naive excited state learning approach, we expect that additional inductive biases—such as employing symmetries to select excited states from a particular Hilbert space sector, tracking level crossings, and adding orthogonality penalties to the loss function—will improve excited state generalization.


\section{VMC comparison}
\label{secVMCComparison}

The supervised results above establish that a single parameter conditioned wave function can learn an accurate map from Hamiltonian parameters to quantum ground states. 

In the following, we perform unsupervised training of our foundation model, where the training is performed through variational Monte Carlo (VMC) energy minimization, without using exact wave functions or exact energies as training data~\cite{CarleoTroyer2017,Sorella2017,Rende2025}.
Specifically, for a collection of Hamiltonians indexed by
$\boldsymbol{\lambda}$, we minimize the ensemble variational energy
\begin{equation}
    \mathcal{L}_{\mathrm{VMC}}(\theta)
    =
    \sum_{\boldsymbol{\lambda}\in\mathcal{S}}
    w_{\boldsymbol{\lambda}}
    E_{\theta}(\boldsymbol{\lambda}),
    \qquad
    \sum_{\boldsymbol{\lambda}\in\mathcal{S}}
    w_{\boldsymbol{\lambda}}=1,
    \label{eqVMCLoss}
\end{equation}
where
\begin{equation}
    E_{\theta}(\boldsymbol{\lambda})
    =
    \frac{
    \langle
    \Psi_{\theta}(\boldsymbol{\lambda})
    |
    \hat H_{\boldsymbol{\lambda}}
    |
    \Psi_{\theta}(\boldsymbol{\lambda})
    \rangle
    }{
    \langle
    \Psi_{\theta}(\boldsymbol{\lambda})
    |
    \Psi_{\theta}(\boldsymbol{\lambda})
    \rangle
    }.
    \label{eqVMCEnergy}
\end{equation}
This expectation value is estimated from configurations sampled according to
\begin{equation}
    p_{\theta}(\mathbf n|\boldsymbol{\lambda})
    =
    \frac{
    |\psi_{\theta}(\mathbf n;\boldsymbol{\lambda})|^2
    }{
    \sum_{\mathbf n'}
    |\psi_{\theta}(\mathbf n';\boldsymbol{\lambda})|^2
    },
\end{equation}
using the local energy
\begin{equation}
    E_{\mathrm{loc}}(\mathbf n;\boldsymbol{\lambda})
    =
    \sum_{\mathbf n'}
    \langle\mathbf n|
    \hat H_{\boldsymbol{\lambda}}
    |\mathbf n'\rangle
    \frac{
    \psi_{\theta}(\mathbf n';\boldsymbol{\lambda})
    }{
    \psi_{\theta}(\mathbf n;\boldsymbol{\lambda})
    }.
    \label{eqLocalEnergy}
\end{equation}

\begin{figure*}
    \centering
    \includegraphics[width=\linewidth]
    {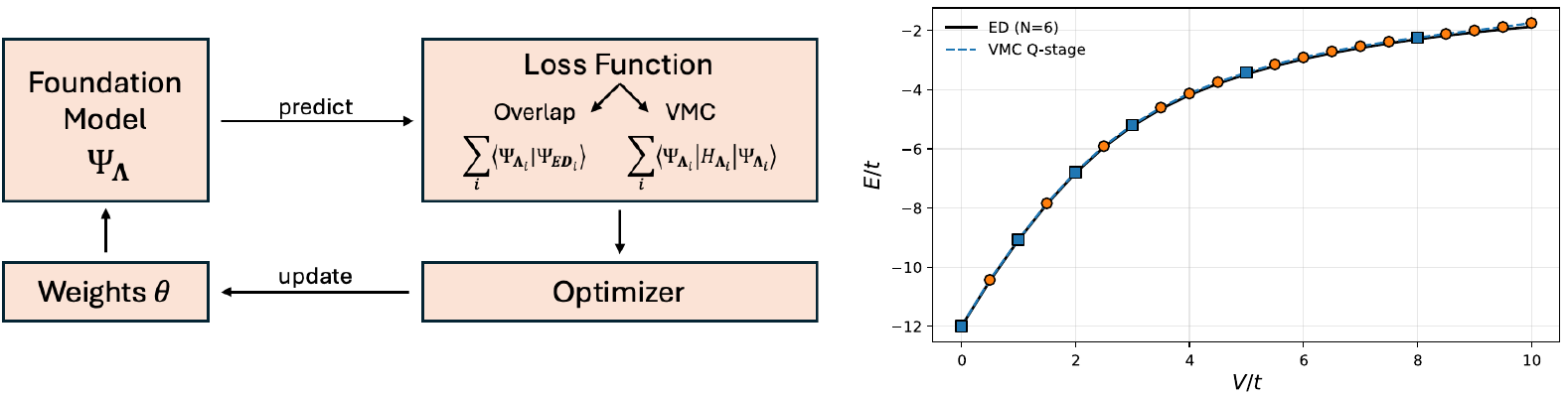}
    \caption{{\textbf{Q-stage workflow and VMC foundation model for $N=6$.} (Left) Full Q-stage workflow with flexible supervised and unsupervised training objective. (Right) VMC foundation model energy comparison for the doped two dimensional square lattice with $N_s=16$ and $N=6$. The black curve gives the ED ground state energy. The blue dashed curve and orange circles give the energy produced by a single Q stage model trained through variational energy minimization. Blue squares indicate the Hamiltonians included in training. The lower panel shows the difference $E_{\theta}-E_{\mathrm{ED}}$. The variational energy closely follows the exact result across the full interaction range, including the Hamiltonians excluded from training.}}
    \label{figVMCEnergy}
\end{figure*}

Figure~\ref{figVMCEnergy} demonstrates that this unsupervised foundation model works in practice, while also demonstrating the flexibility of the architecture and method to be accurate both using overlap and VMC. 
From a sparse set of training Hamiltonians, one VMC energy optimized model accurately reproduces the ground state energy across the full interaction range for particle number $N=6$. Agreement persists at Hamiltonians not included in training, showing that VMC retains the parameter generalization established above through overlap supervision. 
Thus, we explicitly demonstrate the Q-stage framework extends powerfully to VMC energy minimization, allowing the architecture to perform with precise accuracy for both supervised and unsupervised approaches. A more detailed analysis of the overlap versus VMC energy minmization, and the advantages of each objective is provided in the supplementary material~\cite{supplementary}.


The VMC formulation also changes the computational scaling. Exact
diagonalization requires a Hilbert space whose dimension grows
combinatorially with system size. VMC instead evaluates the wave function
only on sampled configurations, so its cost is governed by the number of
samples, the Hamiltonian connectivity, and the neural network evaluation
cost~\cite{Sorella2017}. Neural VMC methods have consequently exhibited
power law scaling in system size
\cite{geier2025self,rende2026scalinglawsneuralnetworkquantum}. 
It can therefore be extended to much larger two dimensional systems by increasing the available sampling and compute. In particular, exact wave functions should be used for overlap training where they are available because they constrain the complete amplitude, phase, and correlation structure. 
Using the initialization of the overlap objective, VMC energy minimization can then extend the learned representation to larger systems where exact data no longer exist. 
Their combination joins information rich supervision with scalable variational learning to create a direct path toward foundation models that can accurately predict quantum states beyond the limits of exact diagonalization and two dimensional DMRG.

\section{Discussion}
\label{sec:disc}

We introduced an attention-based foundation model that learns and predicts many-body wavefunctions directly from basis configurations and Hamiltonian parameters, with no additional input knowledge about the system. 
Provided a smooth mapping from parameter space to the wavefunction space exists, our model yields strong generalization across parameter space, and out-of-sample energies are accurate despite no direct training toward it. 
From the predicted wavefunction, equal-time observables follow from post-processing, and the phase diagram is recovered with high fidelity. Inference is fast: once trained, the model evaluates an entire parameter grid in minutes, in sharp contrast to the exponential scaling of ED. These results provide evidence for the generality and predictive power of our foundation model, so that training on a family of Hamiltonians can yield any accurate wavefunction under a unified representation.
 

\par Most attractively, the generalization of our approach is vast. While the current model is restricted by ED supervision, the framework is immediately extensible along several axes: (i) Scaling to larger systems-. The flexibility of our approach allows us to train on smaller system sizes with ED or DMRG validation, and then perform VMC energy minimization on larger lattices. (ii) Broader Hamiltonians-. Because inputs are tokenized basis states plus a shared parameter token, our architecture transfers to other lattice or spin models (e.g., Hubbard, $J_1$--$J_2$) and to bosons by adjusting the site-occupancy vocabulary without redesigning the network. (iii) Continuum models-. Our general framework can be applied to the orbital basis in real space rather than the occupational basis, where the positions of each particle are sampled from the continuum. {Indeed, we recently extended this general architecture to continuum systems with unprecedented accuracy and scaling success~\cite{zaklama2026large}} (iv) Excited states-. Orthogonality penalties (such as Graham-Schmidt orthogonalization) and symmetry sectoring can be applied to find excited states. We leave these extensions to future works. 

\section{Acknowledgements}

It is our pleasure to thank Max Geier, Zoe Zhu, Filippo Gaggioli, David Lin, and Matthew Burruss for helpful discussions and encouragement. This work was supported by  the NSF through Award No. PHY 2425180. TZ was supported by the MIT Dean of Science Graduate Student Fellowship. LF was supported by a Simons Investigator Award from the Simons Foundation. 

\section{Data Availability}
The data that support the findings of this study are available from the
corresponding author upon request.

\section{Code Availability}
The architecture built in this paper is publicly available \url{https://www.deeppsi.ai/code-repository}, along with other AI tools to simulate quantum many-body systems.

\bibliography{ref}

@article{CarleoTroyer2017,
  title = {Solving the quantum many-body problem with artificial neural networks},
  author = {Carleo, Giuseppe and Troyer, Matthias},
  journal = {Science},
  volume = {355},
  number = {6325},
  pages = {602--606},
  year = {2017},
  month = {Feb},
  publisher = {American Association for the Advancement of Science},
  doi = {10.1126/science.aag2302},
  url = {https://www.science.org/doi/10.1126/science.aag2302}
}

@article{ZhangDiVentra2023,
  title = {Transformer quantum state: A multipurpose model for quantum many-body problems},
  author = {Zhang, Yuan-Hang and Di Ventra, Massimiliano},
  journal = {Phys. Rev. B},
  volume = {107},
  number = {7},
  pages = {075147},
  year = {2023},
  month = {Feb},
  publisher = {American Physical Society},
  doi = {10.1103/PhysRevB.107.075147},
  url = {https://link.aps.org/doi/10.1103/PhysRevB.107.075147}
}

@article{Rende2025,
  title = {Foundation neural-networks quantum states as a unified Ansatz for multiple hamiltonians},
  author = {Rende, Riccardo and Viteritti, Luciano Loris and Becca, Federico and Scardicchio, Antonello and Laio, Alessandro and Carleo, Giuseppe},
  journal = {Nat. Commun.},
  volume = {16},
  number = {1},
  pages = {7213},
  year = {2025},
  month = {Aug},
  publisher = {Nature Publishing Group},
  doi = {10.1038/s41467-025-62098-x},
  url = {https://www.nature.com/articles/s41467-025-62098-x}
}

@article{YangZhang2021,
  title = {Electronic structures, charge transfer, and charge order in twisted transition metal dichalcogenide bilayers},
  author = {Zhang, Yang and Liu, Tongtong and Fu, Liang},
  journal = {Phys. Rev. B},
  volume = {103},
  issue = {15},
  pages = {155142},
  numpages = {6},
  year = {2021},
  month = {Apr},
  publisher = {American Physical Society},
  doi = {10.1103/PhysRevB.103.155142},
  url = {https://link.aps.org/doi/10.1103/PhysRevB.103.155142}
}

@misc{guerci2025ferromagneticsuperconductivityexcitoniccooper,
      title={Ferromagnetic superconductivity with excitonic Cooper pairs: Application to $\Gamma$-valley twisted semiconductors}, 
      author={Daniele Guerci and Liang Fu},
      year={2025},
      eprint={2503.05863},
      archivePrefix={arXiv},
      primaryClass={cond-mat.supr-con},
      url={https://arxiv.org/abs/2503.05863}, 
}

@misc{supplementary,
	Note = {see supplementary Material at url .... for a details on....}}

@misc{chen2025neuralnetworkaugmentedpfaffianwavefunctions,
      title={Neural Network-Augmented Pfaffian Wave-functions for Scalable Simulations of Interacting Fermions}, 
      author={Ao Chen and Zhou-Quan Wan and Anirvan Sengupta and Antoine Georges and Christopher Roth},
      year={2025},
      eprint={2507.10705},
      archivePrefix={arXiv},
      primaryClass={cond-mat.str-el},
      url={https://arxiv.org/abs/2507.10705}, 
}

@article{Smith2024,
  title = {Unified Variational Approach Description of Ground-State Phases of the Two-Dimensional Electron Gas},
  author = {Smith, Conor and Chen, Yixiao and Levy, Ryan and Yang, Yubo and Morales, Miguel A. and Zhang, Shiwei},
  journal = {Phys. Rev. Lett.},
  volume = {133},
  issue = {26},
  pages = {266504},
  numpages = {6},
  year = {2024},
  month = {Dec},
  publisher = {American Physical Society},
  doi = {10.1103/PhysRevLett.133.266504},
  url = {https://link.aps.org/doi/10.1103/PhysRevLett.133.266504}
}

@article{Viteritti2023Transformer,
  title = {Transformer Variational Wave Functions for Frustrated Quantum Spin Systems},
  author = {Viteritti, Luciano Loris and Rende, Riccardo and Becca, Federico},
  journal = {Phys. Rev. Lett.},
  volume = {130},
  number = {23},
  pages = {236401},
  year = {2023},
  month = {Jun},
  publisher = {American Physical Society},
  doi = {10.1103/PhysRevLett.130.236401},
  url = {https://link.aps.org/doi/10.1103/PhysRevLett.130.236401}
}

@article{Psiformer2022,
  title = {A Self-Attention Ansatz for Ab-initio Quantum Chemistry},
  author = {von Glehn, Ingrid and Spencer, James S. and Pfau, David},
  journal = {arXiv},
  eprint = {2211.13672},
  archivePrefix = {arXiv},
  primaryClass = {physics.chem-ph},
  year = {2022},
  url = {https://arxiv.org/abs/2211.13672}
}

@article{Sharir2020,
  title = {Deep Autoregressive Models for the Efficient Variational Simulation of Many-Body Quantum Systems},
  author = {Sharir, Or and Levine, Yoav and Wies, Noam and Carleo, Giuseppe and Shashua, Amnon},
  journal = {Phys. Rev. Lett.},
  volume = {124},
  number = {2},
  pages = {020503},
  year = {2020},
  month = {Jan},
  publisher = {American Physical Society},
  doi = {10.1103/PhysRevLett.124.020503},
  url = {https://link.aps.org/doi/10.1103/PhysRevLett.124.020503}
}

@article{Pfau2020,
  title = {Ab initio solution of the many-electron Schr{\"o}dinger equation with deep neural networks},
  author = {Pfau, David and Spencer, James S. and Matthews, Alexander G. D. G. and Foulkes, W. M. C.},
  journal = {Phys. Rev. Research},
  volume = {2},
  number = {3},
  pages = {033429},
  year = {2020},
  month = {Sep},
  publisher = {American Physical Society},
  doi = {10.1103/PhysRevResearch.2.033429},
  url = {https://link.aps.org/doi/10.1103/PhysRevResearch.2.033429}
}

@misc{Bommasani2021,
  title   = {On the Opportunities and Risks of Foundation Models},
  author  = {Bommasani, Rishi and others},
  year    = {2021},
  eprint  = {2108.07258},
  archivePrefix = {arXiv},
  primaryClass  = {cs.LG},
  url     = {https://arxiv.org/abs/2108.07258},
  note    = {CRFM Report, Stanford}
}

@inproceedings{Vaswani2017,
  title     = {Attention Is All You Need},
  author    = {Vaswani, Ashish and Shazeer, Noam and Parmar, Niki and Uszkoreit, Jakob and Jones, Llion and Gomez, Aidan N. and Kaiser, Lukasz and Polosukhin, Illia},
  booktitle = {Advances in Neural Information Processing Systems (NeurIPS)},
  year      = {2017},
  url       = {https://papers.neurips.cc/paper/7181-attention-is-all-you-need.pdf}
}

@inproceedings{Brown2020,
  title     = {Language Models are Few-Shot Learners},
  author    = {Brown, Tom B. and Mann, Benjamin and Ryder, Nick and Subbiah, Melanie and Kaplan, Jared and Dhariwal, Prafulla and Neelakantan, Arvind and Shyam, Pranav and Sastry, Girish and Askell, Amanda and Agarwal, Sandhini and Herbert-Voss, Ariel and Krueger, Gretchen and Henighan, Tom and Child, Rewon and Ramesh, Aditya and Ziegler, Daniel M. and Wu, Jeffrey and Winter, Clemens and Hesse, Christopher and Chen, Mark and Sigler, Eric and Litwin, Mateusz and Gray, Scott and Chess, Benjamin and Clark, Jack and Berner, Christopher and McCandlish, Sam and Radford, Alec and Sutskever, Ilya and Amodei, Dario},
  booktitle = {Advances in Neural Information Processing Systems (NeurIPS)},
  year      = {2020},
  url       = {https://proceedings.neurips.cc/paper/2020/file/1457c0d6bfcb4967418bfb8ac142f64a-Paper.pdf}
}

@inproceedings{Radford2021,
  title     = {Learning Transferable Visual Models From Natural Language Supervision},
  author    = {Radford, Alec and Kim, Jong Wook and Hallacy, Chris and Ramesh, Aditya and Goh, Gabriel and Agarwal, Sandhini and Sastry, Girish and Askell, Amanda and Mishkin, Pamela and Clark, Jack and Krueger, Gretchen and Sutskever, Ilya},
  booktitle = {Proceedings of the 38th International Conference on Machine Learning (ICML)},
  series    = {Proceedings of Machine Learning Research},
  volume    = {139},
  pages     = {8748--8763},
  year      = {2021},
  url       = {https://proceedings.mlr.press/v139/radford21a/radford21a.pdf}
}

@inproceedings{Ramesh2021,
  title     = {Zero-Shot Text-to-Image Generation},
  author    = {Ramesh, Aditya and Pavlov, Mikhail and Goh, Gabriel and Gray, Scott and Voss, Chelsea and Radford, Alec and Chen, Mark and Sutskever, Ilya},
  booktitle = {Proceedings of the 38th International Conference on Machine Learning (ICML)},
  series    = {Proceedings of Machine Learning Research},
  volume    = {139},
  pages     = {8821--8831},
  year      = {2021},
  url       = {https://proceedings.mlr.press/v139/ramesh21a.html}
}

@article{Jumper2021,
  title   = {Highly accurate protein structure prediction with AlphaFold},
  author  = {Jumper, John and Evans, Richard and Pritzel, Alexander and Green, Tim and Figurnov, Michael and Ronneberger, Olaf and Tunyasuvunakool, Kathryn and Bates, Russell and {\v{Z}}{\'\i}dek, Anna and Potapenko, Anna and others},
  journal = {Nature},
  volume  = {596},
  number  = {7873},
  pages   = {583--589},
  year    = {2021},
  doi     = {10.1038/s41586-021-03819-2},
  url     = {https://www.nature.com/articles/s41586-021-03819-2}
}

@misc{Evans2022,
  title   = {Protein complex prediction with {AlphaFold}-Multimer},
  author  = {Evans, Richard and O'Neill, Michael and Pritzel, Alexander and Antropova, Natasha and Senior, Andrew and Green, Tim and {\v{Z}}{\'\i}dek, Anna and Bates, Russell and Blackwell, Sam and Yim, Jason and others},
  year    = {2022},
  eprint  = {2021.10.04.463034},
  archivePrefix = {bioRxiv},
  doi     = {10.1101/2021.10.04.463034},
  url     = {https://www.biorxiv.org/content/10.1101/2021.10.04.463034v2}
}

@article{TQS2023,
  title   = {Transformer quantum state: A multipurpose model for quantum many-body problems},
  author  = {Zhang, Yuan-Hang and Di Ventra, Massimiliano},
  journal = {Phys. Rev. B},
  volume  = {107},
  number  = {7},
  pages   = {075147},
  year    = {2023},
  doi     = {10.1103/PhysRevB.107.075147},
  url     = {https://link.aps.org/doi/10.1103/PhysRevB.107.075147}
}

@book{Sorella2017,
  title     = {Quantum Monte Carlo Approaches for Correlated Systems},
  author    = {Becca, Federico and Sorella, Sandro},
  publisher = {Cambridge University Press},
  address   = {Cambridge, UK},
  year      = {2017},
  doi       = {10.1017/9781316417041},
  isbn      = {9781107129931},
  url       = {https://www.cambridge.org/core/books/quantum-monte-carlo-approaches-for-correlated-systems/EB88C86BD9553A0738BDAE400D0B2900}
}

@article{Haldane1982,
  title = {Spontaneous dimerization in the $S=\frac{1}{2}$ Heisenberg antiferromagnetic chain with competing interactions},
  author = {Haldane, F. D. M.},
  journal = {Phys. Rev. B},
  volume = {25},
  number = {7},
  pages = {4925--4928},
  year = {1982},
  month = {Apr},
  publisher = {American Physical Society},
  doi = {10.1103/PhysRevB.25.4925},
  url = {https://link.aps.org/doi/10.1103/PhysRevB.25.4925}
}

@article{Carrasquilla_2017,
   title={Machine learning phases of matter},
   volume={13},
   ISSN={1745-2481},
   url={http://dx.doi.org/10.1038/nphys4035},
   DOI={10.1038/nphys4035},
   number={5},
   journal={Nature Physics},
   publisher={Springer Science and Business Media LLC},
   author={Carrasquilla, Juan and Melko, Roger G.},
   year={2017},
   month=feb, pages={431–434} }

@article{Wang_2014,
   title={Fermionic quantum critical point of spinless fermions on a honeycomb lattice},
   volume={16},
   ISSN={1367-2630},
   url={http://dx.doi.org/10.1088/1367-2630/16/10/103008},
   DOI={10.1088/1367-2630/16/10/103008},
   number={10},
   journal={New Journal of Physics},
   publisher={IOP Publishing},
   author={Wang, Lei and Corboz, Philippe and Troyer, Matthias},
   year={2014},
   month=oct, pages={103008} }

@article{Mila_1993,
   title={Phase Diagram of the One-Dimensional Extended Hubbard Model at Quarter-Filling},
   volume={24},
   ISSN={1286-4854},
   url={http://dx.doi.org/10.1209/0295-5075/24/2/010},
   DOI={10.1209/0295-5075/24/2/010},
   number={2},
   journal={Europhysics Letters (EPL)},
   publisher={IOP Publishing},
   author={Mila, F and Zotos, X},
   year={1993},
   month=oct, pages={133–138} }

@article{Stokes2020,
  title = {Phases of two-dimensional spinless lattice fermions with first-quantized deep neural-network quantum states},
  author = {Stokes, James and Moreno, Javier Robledo and Pnevmatikakis, Eftychios A. and Carleo, Giuseppe},
  journal = {Phys. Rev. B},
  volume = {102},
  issue = {20},
  pages = {205122},
  numpages = {10},
  year = {2020},
  month = {Nov},
  publisher = {American Physical Society},
  doi = {10.1103/PhysRevB.102.205122},
  url = {https://link.aps.org/doi/10.1103/PhysRevB.102.205122}
}

@article{Gubernatis1985,
  title   = {Two-dimensional spin-polarized fermion lattice gases},
  author  = {Gubernatis, J. E. and Scalapino, D. J. and Sugar, R. L. and Toussaint, W. D.},
  journal = {Phys. Rev. B},
  volume  = {32},
  pages   = {103--112},
  year    = {1985},
  month   = {Jul},
  publisher = {American Physical Society},
  doi     = {10.1103/PhysRevB.32.103},
  url     = {https://link.aps.org/doi/10.1103/PhysRevB.32.103}
}

@article{Czart2008,
  title   = {Charge ordering and phase separations in the spinless fermion model with repulsive intersite interaction},
  author  = {Czart, W. R. and Robaszkiewicz, S. and Tobijaszewska, B.},
  journal = {Acta Phys. Pol. A},
  volume  = {114},
  pages   = {129--136},
  year    = {2008}
}

@article{deWoul2010,
  title   = {Partially gapped fermions in 2D},
  author  = {de Woul, Jonas and Langmann, Edwin},
  journal = {J. Stat. Phys.},
  volume  = {139},
  pages   = {1033--1067},
  year    = {2010}
}

@article{Song2014,
  title   = {Monte Carlo simulations of two-dimensional fermion systems with string-bond states},
  author  = {Song, J.-P. and Clay, R. T.},
  journal = {Phys. Rev. B},
  volume  = {89},
  pages   = {075101},
  year    = {2014},
  month   = {Feb},
  publisher = {American Physical Society},
  doi     = {10.1103/PhysRevB.89.075101},
  url     = {https://link.aps.org/doi/10.1103/PhysRevB.89.075101}
}

@article{Sikora2015,
  title   = {Variational Monte Carlo simulations using tensor-product projected states},
  author  = {Sikora, O. and Chang, H.-W. and Chou, C.-P. and Pollmann, F. and Kao, Y.-J.},
  journal = {Phys. Rev. B},
  volume  = {91},
  pages   = {165113},
  year    = {2015},
  month   = {Apr},
  publisher = {American Physical Society},
  doi     = {10.1103/PhysRevB.91.165113},
  url     = {https://link.aps.org/doi/10.1103/PhysRevB.91.165113}
}

@article{Viteritti_2025,
   title={Transformer wave function for two dimensional frustrated magnets: Emergence of a spin-liquid phase in the Shastry-Sutherland model},
   volume={111},
   ISSN={2469-9969},
   url={http://dx.doi.org/10.1103/PhysRevB.111.134411},
   DOI={10.1103/physrevb.111.134411},
   number={13},
   journal={Physical Review B},
   publisher={American Physical Society (APS)},
   author={Viteritti, Luciano Loris and Rende, Riccardo and Parola, Alberto and Goldt, Sebastian and Becca, Federico},
   year={2025},
   month=apr }

@article{geier2025self,
  title = {Self-attention neural network for solving correlated electron problems in solids},
  author = {Geier, Max and Nazaryan, Khachatur and Zaklama, Timothy and Fu, Liang},
  journal = {Phys. Rev. B},
  volume = {112},
  issue = {4},
  pages = {045119},
  numpages = {16},
  year = {2025},
  month = {Jul},
  publisher = {American Physical Society},
  doi = {10.1103/qxc3-bkc7},
  url = {https://link.aps.org/doi/10.1103/qxc3-bkc7}
}

@article{teng2025solving,
  title = {Solving the fractional quantum Hall problem with self-attention neural network},
  author = {Teng, Yi and Dai, David D. and Fu, Liang},
  journal = {Phys. Rev. B},
  volume = {111},
  issue = {20},
  pages = {205117},
  numpages = {10},
  year = {2025},
  month = {May},
  publisher = {American Physical Society},
  doi = {10.1103/PhysRevB.111.205117},
  url = {https://link.aps.org/doi/10.1103/PhysRevB.111.205117}
}

@article{nazaryan2025finding,
  title={Artificial Intelligence for Quantum Matter: Finding a Needle in a Haystack},
  author={Nazaryan, Khachatur and Gaggioli, Filippo and Fu, Liang},
  journal={arXiv preprint arXiv:2507.13322},
  year={2025}
}

@article{luo2019backflow,
  title = {Backflow Transformations via Neural Networks for Quantum Many-Body Wave Functions},
  author = {Luo, Di and Clark, Bryan K.},
  journal = {Phys. Rev. Lett.},
  volume = {122},
  issue = {22},
  pages = {226401},
  numpages = {6},
  year = {2019},
  month = {Jun},
  publisher = {American Physical Society},
  doi = {10.1103/PhysRevLett.122.226401},
  url = {https://link.aps.org/doi/10.1103/PhysRevLett.122.226401}
}

@article{foster2025ab,
  title={An ab initio foundation model of wavefunctions that
    accurately describes chemical bond breaking},
  author={Foster, Adam and Schatzle, Zeno and Szabo, P. Bernat and Cheng, Lixue and Kohler, Jonas and Cassalla, Gino and Gao, Nicholas and Li, Jiawei and Noe, Frank and Hermann, Jan},
  journal={arXiv preprint arXiv:2506.19960},
  year={2025}
}

@article{zhu2023hubbard,
  title = {HubbardNet: Efficient predictions of the Bose-Hubbard model spectrum with deep neural networks},
  author = {Zhu, Ziyan and Mattheakis, Marios and Pan, Weiwei and Kaxiras, Efthimios},
  journal = {Phys. Rev. Res.},
  volume = {5},
  issue = {4},
  pages = {043084},
  numpages = {11},
  year = {2023},
  month = {Oct},
  publisher = {American Physical Society},
  doi = {10.1103/PhysRevResearch.5.043084},
  url = {https://link.aps.org/doi/10.1103/PhysRevResearch.5.043084}
}

@inproceedings{He2016DeepResNet,
  title     = {Deep Residual Learning for Image Recognition},
  author    = {He, Kaiming and Zhang, Xiangyu and Ren, Shaoqing and Sun, Jian},
  booktitle = {Proceedings of the IEEE Conference on Computer Vision and Pattern Recognition (CVPR)},
  year      = {2016},
  pages     = {770--778},
  publisher = {IEEE},
  doi       = {10.1109/CVPR.2016.90},
  url       = {https://doi.org/10.1109/CVPR.2016.90}
}

@article{Ba2016LayerNorm,
  title         = {Layer Normalization},
  author        = {Ba, Jimmy Lei and Kiros, Jamie Ryan and Hinton, Geoffrey E.},
  journal       = {arXiv preprint arXiv:1607.06450},
  year          = {2016},
  archivePrefix = {arXiv},
  eprint        = {1607.06450},
  primaryClass  = {cs.LG},
  url           = {https://arxiv.org/abs/1607.06450}
}

@article{Slagle2020charge,
   title={Charge transfer excitations, pair density waves, and superconductivity in moiré materials},
   volume={102},
   ISSN={2469-9969},
   url={http://dx.doi.org/10.1103/PhysRevB.102.235423},
   DOI={10.1103/physrevb.102.235423},
   number={23},
   journal={Physical Review B},
   publisher={American Physical Society (APS)},
   author={Slagle, Kevin and Fu, Liang},
   year={2020},
   month=dec }

@article{
crepel2021new,
author = {Valentin Crépel  and Liang Fu},
title = {New mechanism and exact theory of superconductivity from strong repulsive interaction},
journal = {Science Advances},
volume = {7},
number = {30},
pages = {eabh2233},
year = {2021},
doi = {10.1126/sciadv.abh2233},
URL = {https://www.science.org/doi/abs/10.1126/sciadv.abh2233},
eprint = {https://www.science.org/doi/pdf/10.1126/sciadv.abh2233}}

@article{He2023superconductivity,
  title = {Superconductivity of repulsive spinless fermions with sublattice potentials},
  author = {He, Yuchi and Yang, Kang and Profe, Jonas B. and Bergholtz, Emil J. and Kennes, Dante M.},
  journal = {Phys. Rev. Res.},
  volume = {5},
  issue = {1},
  pages = {L012009},
  numpages = {7},
  year = {2023},
  month = {Jan},
  publisher = {American Physical Society},
  doi = {10.1103/PhysRevResearch.5.L012009},
  url = {https://link.aps.org/doi/10.1103/PhysRevResearch.5.L012009}
}

@article{Liang2018solving,
  title = {Solving frustrated quantum many-particle models with convolutional neural networks},
  author = {Liang, Xiao and Liu, Wen-Yuan and Lin, Pei-Ze and Guo, Guang-Can and Zhang, Yong-Sheng and He, Lixin},
  journal = {Phys. Rev. B},
  volume = {98},
  issue = {10},
  pages = {104426},
  numpages = {6},
  year = {2018},
  month = {Sep},
  publisher = {American Physical Society},
  doi = {10.1103/PhysRevB.98.104426},
  url = {https://link.aps.org/doi/10.1103/PhysRevB.98.104426}
}

@misc{gaggioli2025electronic,
      title={Electronic crystals and quasicrystals in semiconductor quantum wells: an AI-powered discovery}, 
      author={Filippo Gaggioli and Pierre-Antoine Graham and Liang Fu},
      year={2025},
      eprint={2512.10909},
      archivePrefix={arXiv},
      primaryClass={cond-mat.str-el},
      url={https://arxiv.org/abs/2512.10909}, 
}

@misc{zaklama2026large,
      title={Large Electron Model: A Universal Ground State Predictor}, 
      author={Timothy Zaklama and Max Geier and Liang Fu},
      year={2026},
      eprint={2603.02346},
      archivePrefix={arXiv},
      primaryClass={cond-mat.str-el},
      url={https://arxiv.org/abs/2603.02346}, 
}

@misc{rende2026scalinglawsneuralnetworkquantum,
      title={Scaling Laws for Neural-Network Quantum States}, 
      author={Riccardo Rende and Alessandro Sinibaldi and Luciano Loris Viteritti and Roeland Wiersema and Antoine Georges and Giuseppe Carleo},
      year={2026},
      eprint={2606.02794},
      archivePrefix={arXiv},
      primaryClass={cond-mat.dis-nn},
      url={https://arxiv.org/abs/2606.02794}, 
}


\onecolumngrid
\newpage
\makeatletter 

\begin{center}
\textbf{\large Supplementary Material for \emph{Attention-Based Foundation Model for Quantum States}} \\[10pt]
Timothy Zaklama$^{1}$, Daniele Guerci$^{1}$, and Liang Fu$^1$ \\
\textit{$^1$Department of Physics, Massachusetts Institute of Technology, Cambridge, MA-02139, USA}\\
\end{center}
\vspace{10pt}

\setcounter{figure}{0}
\setcounter{section}{0}
\setcounter{equation}{0}

\renewcommand{\thefigure}{S\@arabic\c@figure}
\makeatother

\appendix

\begin{center}
    \textbf{CONTENTS}
\end{center}

I. Order parameter calculation \hfill 2

II. Energy minimization \hfill 3

III. VMC Comparison \hfill 4

IV. Ablations \hfill 7

\quad A. Why cross-attention is used once after the first SA layer \hfill 8

\quad B. Parameter-aware pooling (PAP) \hfill 8

V. Resolution of residual degeneracies \hfill 9

VI. Extended data \hfill 9

\quad A. Energy landscape over the full $(V/t,N)$ grid \hfill 9

\quad B. Pair binding energy across $(V/t,N)$ \hfill 9

\quad C. Sanity checks \hfill 10

VII. Fock basis states \hfill 10

\setcounter{equation}{0}
\setcounter{figure}{0}
\setcounter{table}{0}
\setcounter{page}{1}
\makeatletter
\renewcommand{\theequation}{S\arabic{equation}}
\renewcommand{\thefigure}{S\arabic{figure}}
\renewcommand{\citenumfont}[1]{S#1}
\renewcommand{\citenumfont}[1]{\textit{#1}}
\begin{widetext}

\section{I. ORDER PARAMETER CALCULATION}

\begin{figure}
    \centering
    \includegraphics[width=\linewidth]{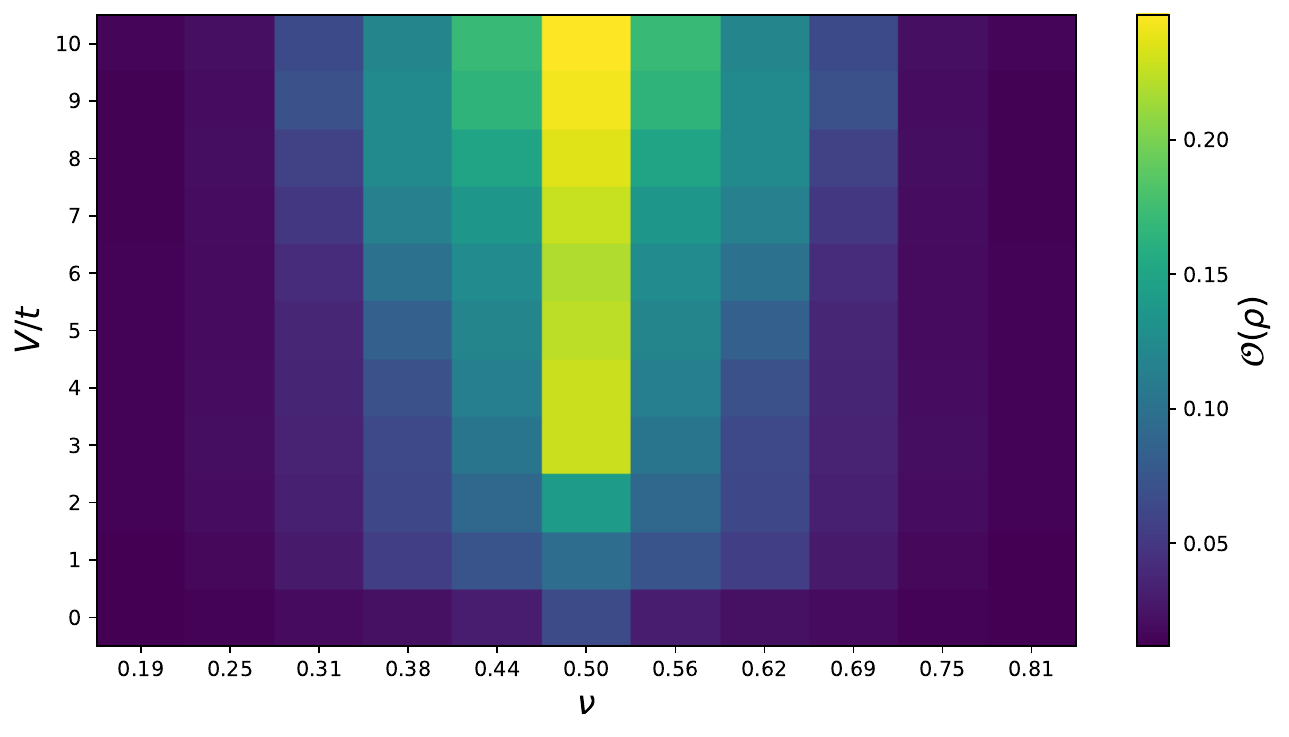}
    \caption{\textbf{Phase diagram of 2D $t$--$V$ model ($L=16$).} {Our phase diagram faithfully reproduces the previous results on this model \cite{Stokes2020}.}}
    \label{fig:heatmap}
\end{figure}

Directly applying the generalization power of our model to predict a family ground states, we build the full phase diagram for the 2 dimensional square lattice by calculating the checkerboard charge density correlation function:

\begin{equation}\label{square_order}
    \begin{split}
    &\mathcal{O}(\rho)=\frac{1}{L^2}\sum_{ij}e^{i\mathbf Q\cdot(\mathbf R_i+\mathbf R_j)} (G_{ij}-\nu^2), \\ &G_{ij} = \langle\Psi_\theta | \rho_i \rho_j | \Psi_\theta \rangle, 
\end{split}
\end{equation}
for total number of sites $L$ and density-density correlation function $G_{ij}$. In Eq.~\eqref{square_order}, $\mathbf{Q} = (\pi, \pi)/a$ defines the modulation of the charge density wave pattern, while $\mathbf{R}_i = n\mathbf{a}_1 + m\mathbf{a}_2$ denotes the position of site $i$ on the two-dimensional lattice.
The quantity, $\mathcal{O}(\rho)$, is the square of the conventional order parameter~\cite{Stokes2020}, and measures the ground state’s tendency toward checkerboard charge ordering, reaching $1/4$ when fully polarized.
We expect this order parameter to increase with $V/t$ and decrease with $N$, creating a quadratic phase transition boundary in the $(V/t,N)$ plane. 
We train on ten $(V/t,N)$ points and predict the entire $(V/t,N)$ plane (over one hundred Hamiltonians) with single training run (Fig.~\ref{fig:heatmap}). The resulting phase diagram faithfully reproduces the metallic regime at small $V/t$ and the checkerboard charge order near half filling at larger $V/t$. 
Amazingly, we did not apply any ground state degenerate selection, and naively selected the ground state based on the lowest energy value. 
Thus the phase boundary agrees with the most recent results on the model~\cite{Stokes2020}, but without imposing symmetries, hand-designed features, or even Hilbert-space sectoring (biasing the trained wavefunctions to a specific symmetry sector). 
To our knowledge, this is the first demonstration of full $(V/t,N)$ phase-diagram reconstruction from sparse wavefunction supervision in a minimal-bias transformer without sectoring.

\section{II. ENERGY MINIMIZATION}
\label{sec:energy_from_overlap}

Our training objective is to maximize overlap (fidelity) with exact ground states across a distribution of Hamiltonian parameters, 
\begin{align}
\mathcal L(\theta) 
&= 1 - \sum_{\boldsymbol\lambda\in \mathcal S} w_{\boldsymbol\lambda}\,
\mathcal O_v(\boldsymbol\lambda),\qquad
\sum_{\boldsymbol\lambda\in\mathcal S} w_{\boldsymbol\lambda}=1,
\label{eq:loss-disc_supp}
\end{align}
with per–instance fidelity
\begin{equation}
\mathcal O_v(\boldsymbol\lambda) 
= \frac{\bigl|\langle{\Psi_\theta(\boldsymbol\lambda)}|{\Psi_{\mathrm{ED}}(\boldsymbol\lambda)}\rangle\bigr|^2}{\sqrt{\|\Psi_\theta(\boldsymbol\lambda)\|^2\,\|\Psi_{\mathrm{ED}}(\boldsymbol\lambda)\|^2}}.
\label{eq:overlap_supp}
\end{equation}
After training we evaluate the variational energy via the Rayleigh quotient,
\begin{equation}
E_\theta(\boldsymbol\lambda)
=\frac{\langle \Psi_\theta(\boldsymbol\lambda)|H(\boldsymbol\lambda)|\Psi_\theta(\boldsymbol\lambda)\rangle}
      {\langle \Psi_\theta(\boldsymbol\lambda)|\Psi_\theta(\boldsymbol\lambda)\rangle}.
\label{eq:rayleigh}
\end{equation}

Let $\{|{\phi_n}\rangle\}$ be the ED eigenbasis with $H|{\phi_n}\rangle=E_n|{\phi_n}\rangle$ and $E_{ED}\le E_1\le\cdots$. 
Writing the normalized variational state as $|{\psi_\theta}\rangle=\sum_n c_n|{\phi_n}\rangle$ gives the exact identity
\begin{equation}
E_\theta - E_{ED}=\sum_{n>0} |c_n|^2\,(E_n-E_{ED}).
\label{eq:energy-exact}
\end{equation}
Denoting the fidelity by $F=|\langle\phi_0|\psi_\theta\rangle|^2=|c_0|^2$ and the residual weight by $\eta=1-F=\sum_{n>0}|c_n|^2$, first–order perturbation theory and Eq.~\eqref{eq:energy-exact} imply the two–sided bound
\begin{equation}
(E_1-E_{ED})\,\eta \ \le\ E_\theta-E_{ED}\ \le\ (E_{\max}-E_{ED})\,\eta,
\label{eq:band-bounds}
\end{equation}
where $E_{\max}$ is the top of the many–body spectrum in the symmetry sector. 
Thus small infidelity does not necessarily imply small energy error: if the residual weight $\eta$ lands on high–lying excitations, the upper bound may be nearly saturated. 
This effect is amplified in the $t$–$V$ model as $V/t$ grows, where the spectral width scales with the interaction energy accumulated over $O(L^2)$ bonds, so even $\eta\sim 10^{-3}$ can yield an absolute error of order $10^{-1}$. 
The phenomenon is clearly visible in Fig.~\ref{fig:overlapvsEnergy}: the perpendicular component $1-\mathcal O_v$ shrinks with $V/t$, yet the energy error plateaus and can even increase at large $V/t$ because the many-body energy width grows while $|E_{ED}|$ decreases, inflating the relative error.

A complementary indicator is the energy variance
\begin{equation}
\mathrm{Var}_\theta(H)=
\langle H^2\rangle_\theta-\langle H\rangle_\theta^2
=\sum_{n>0}|c_n|^2\,(E_n-E_\theta)^2,
\label{eq:variance}
\end{equation}
which vanishes if $|{\psi_\theta}\rangle$ is an eigenstate. 
In practice, we observe that instances with comparable fidelity can display markedly different $\mathrm{Var}_\theta(H)$ depending on how the residual weight is distributed across the spectrum; these are precisely the cases where $E_\theta-E_{ED}$ deviates most from the naive expectation set by the fidelity alone.

To reduce the sensitivity to spectral scale, we augment the fidelity objective with an energy–aware term evaluated exactly (no sampling) on the ED sectors:
\begin{equation}
\mathcal L_{\mathrm{tot}}(\theta)=
\underbrace{1-\sum_{\boldsymbol\lambda\in\mathcal S} w_{\boldsymbol\lambda}\,\mathcal O_v(\boldsymbol\lambda)}_{\text{fidelity}}
\;+\;
\alpha\sum_{\boldsymbol\lambda\in\mathcal S} w_{\boldsymbol\lambda}\,
\frac{\bigl(E_\theta(\boldsymbol\lambda)-E_{ED}(\boldsymbol\lambda)\bigr)^2}{(E_{\rm max}(\boldsymbol\lambda)-E_{ED}(\boldsymbol\lambda))^2},
\label{eq:hybrid-loss}
\end{equation}
where $E_{ED}$ is the ED ground energy and the energy deviation $E_\theta(\boldsymbol\lambda)-E_{ED}(\boldsymbol\lambda)$ is measured with respect to the many-body energy width $E_{\max}-E_{ED}$.
An alternative that does not require $E_{ED}$ is variance minimization, replacing the second term by $\beta\,\sum w_{\boldsymbol\lambda}\,\mathrm{Var}_\theta(H(\boldsymbol\lambda))$, cf.~\eqref{eq:variance}. 
Both choices directly penalize the spectral leakage that overlap training is agnostic to.

Empirically, on the $4\times4$ square benchmark we find that a small energy term ($\alpha\!\sim\!10^{-3}$–$10^{-2}$) significantly lowers $E_\theta-E_{ED}$ in the high–$V/t$ regime where the many-body energy width is largest, consistent with Eq.~\eqref{eq:band-bounds}. 
However, at fixed $\alpha$ the overall reduction in mean relative error is smaller than what is achieved by further improving the average fidelity (i.e.\ driving $\eta$ down uniformly across $\boldsymbol\lambda$). 
In other words, raising $F$ globally remains the most effective route to reduce \emph{average} energy error, while an energy–aware term is valuable to suppress the few ``costly'' directions in the perpendicular subspace that otherwise dominate $E_\theta-E_{ED}$ at large spectral scale. 
When the sole goal is energy (or another observable), the network---initialized from our overlap–trained parameters---can be rapidly fine–tuned by minimizing either the Rayleigh quotient $E_\theta(\boldsymbol\lambda)$ or the variance $\mathrm{Var}_\theta(H)$ over the target subset of $\boldsymbol\lambda$; in practice we observe fast convergence under such observable–only refinement. Thus, our representation makes explicit that the non–monotonic behavior in Fig.~\ref{fig:overlapvsEnergy} stems from variations in spectral scale rather than a failure of the overlap objective.

\begin{figure}
    \centering
    \includegraphics[width=\linewidth]{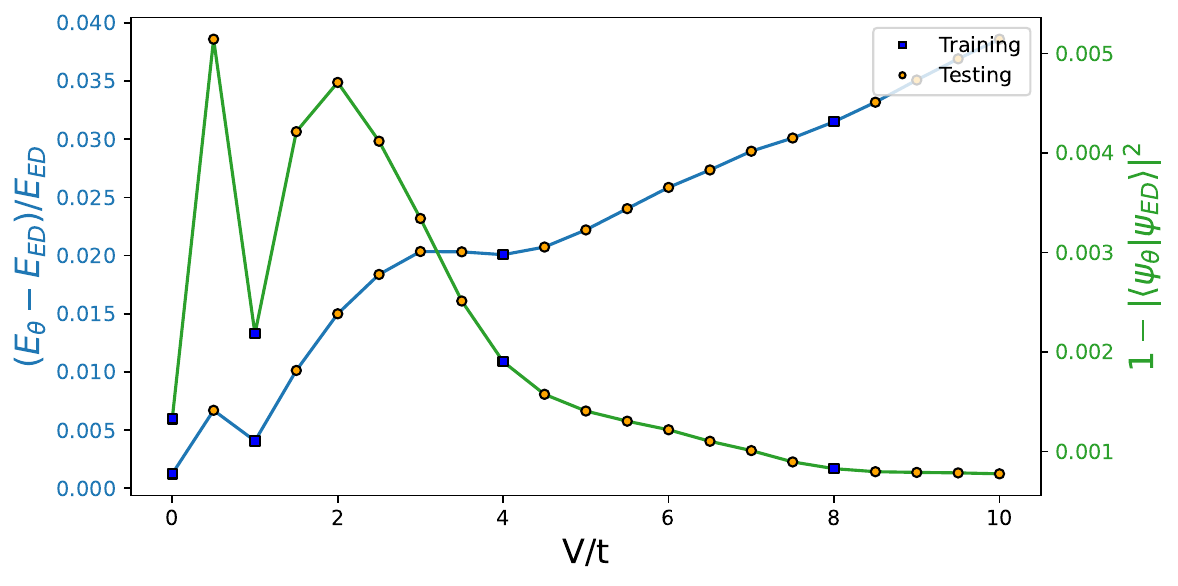}
    \caption{\textbf{Relative energy error and fidelity plotted as a function of parameters $V/t$.} {We can see that higher overlap does not mean lower energy across the parameter space. This means that while improving the overlap does improve variational energy, the energy can still deviate more than expected since the network is agnostic to energy scale. Adding energy expectation to the loss function would diminish this effect.}}
    \label{fig:overlapvsEnergy}
\end{figure}


\section{III. VMC Comparison}
\label{secVMCOverlap}

The results in the main text establish that the foundation model can be
trained without exact wave function data by minimizing the variational
energy over an ensemble of Hamiltonians. Here we examine this approach in
greater detail by comparing the energy and wave function overlap at two
particle numbers. We consider the half filled system with $N=8$, and doped system with $N=6$, as presented
in the main text. The latter provides a
particularly clear demonstration that accurate variational energies do not
require a large overlap with one particular exact eigenvector.

\begin{figure}[t]
\centering
\includegraphics[width=0.78\linewidth]
{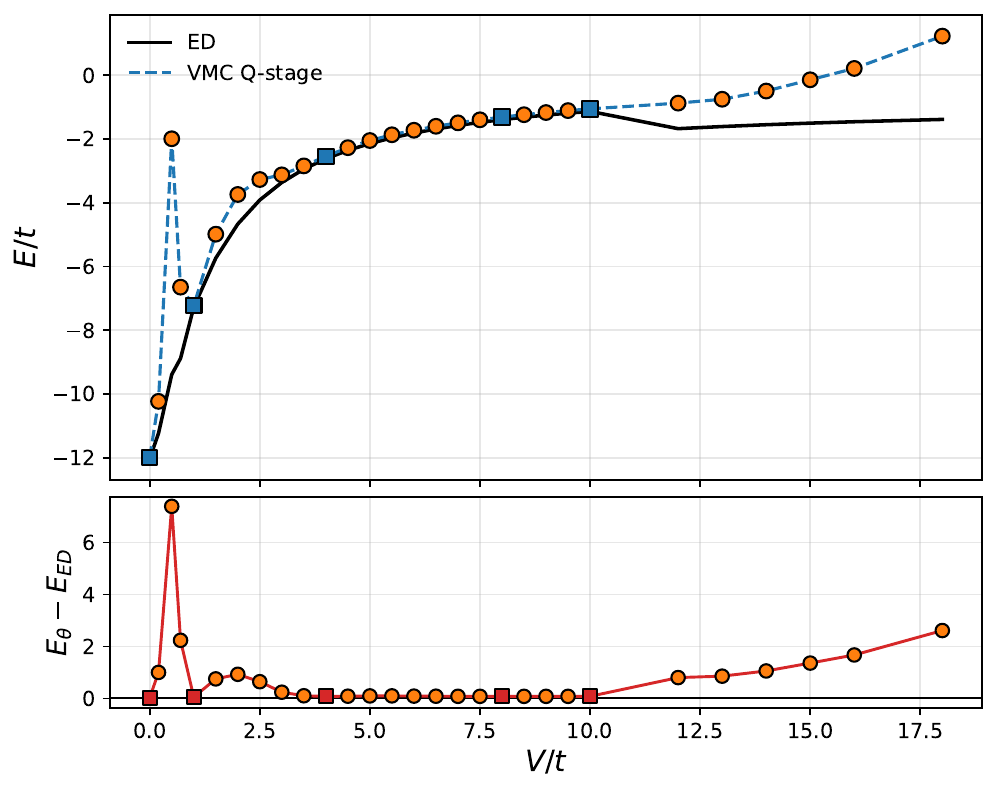}
\caption{\textbf{Variational Monte Carlo energy comparison for the two
    dimensional square lattice with $N_s=16$ and $N=8$.}
    The black curve gives the ED ground state energy. The blue dashed curve
    and orange circles give the energies predicted by a single Q stage model
    trained through VMC energy minimization. Blue squares identify the
    training Hamiltonians. The lower panel gives the energy difference
    $E_{\theta}-E_{\mathrm{ED}}$. The model accurately reproduces the
    training energies and captures the energy throughout much of the
    interpolation region. Larger deviations occur between the weak coupling
    training points and when extrapolating beyond the largest trained
    interaction.}
\label{figN8VMCEnergy}
\end{figure}

Figure~\ref{figN8VMCEnergy} shows the variational energy for $N=8$ over the
range $0\leq V/t\leq18$. The Q-stage result is excellent for medium coupling and large coupling around training points, but deviates for weak coupling around the phase transition. The out of sample energy accuracy obtained from VMC is lower
than that obtained from overlap training with the same sparse training set.
The difference reflects the information contained in the two objectives.
Overlap training constrains the complete complex wave function and therefore
provides coefficient level information throughout the Hilbert space. VMC
instead constrains a stochastic energy expectation and must infer the
remaining wave function structure indirectly. The overlap objective
therefore offers stronger supervision when exact states are available,
whereas the VMC objective removes the system size restriction and directly
targets the variational energy.
\begin{figure}[t]
\centering
\includegraphics[width=\linewidth]
{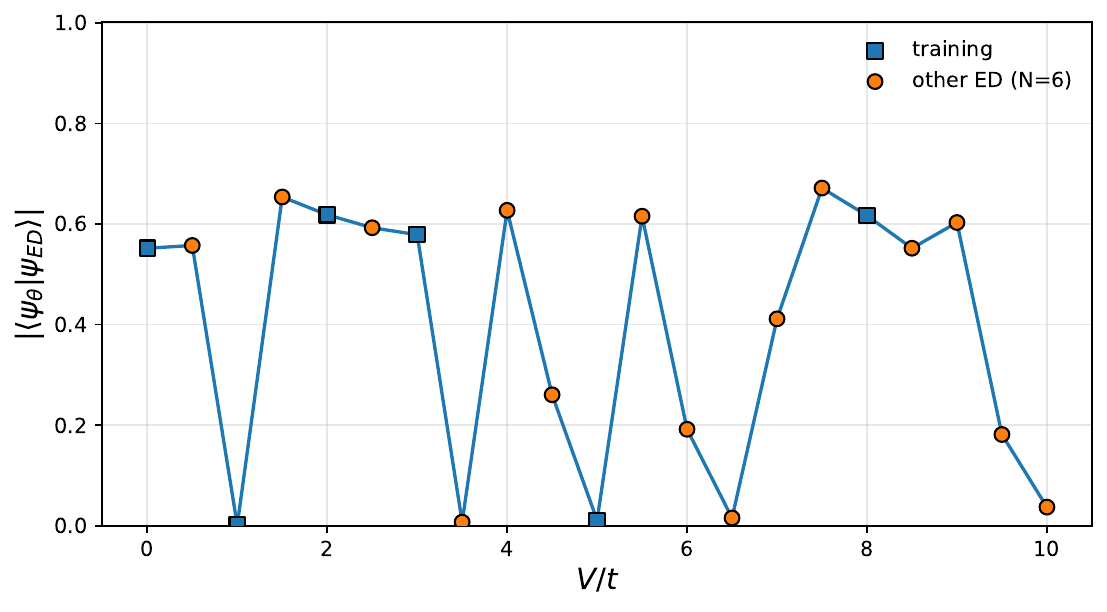}

```
\vspace{0.15cm}

\includegraphics[width=\linewidth]
{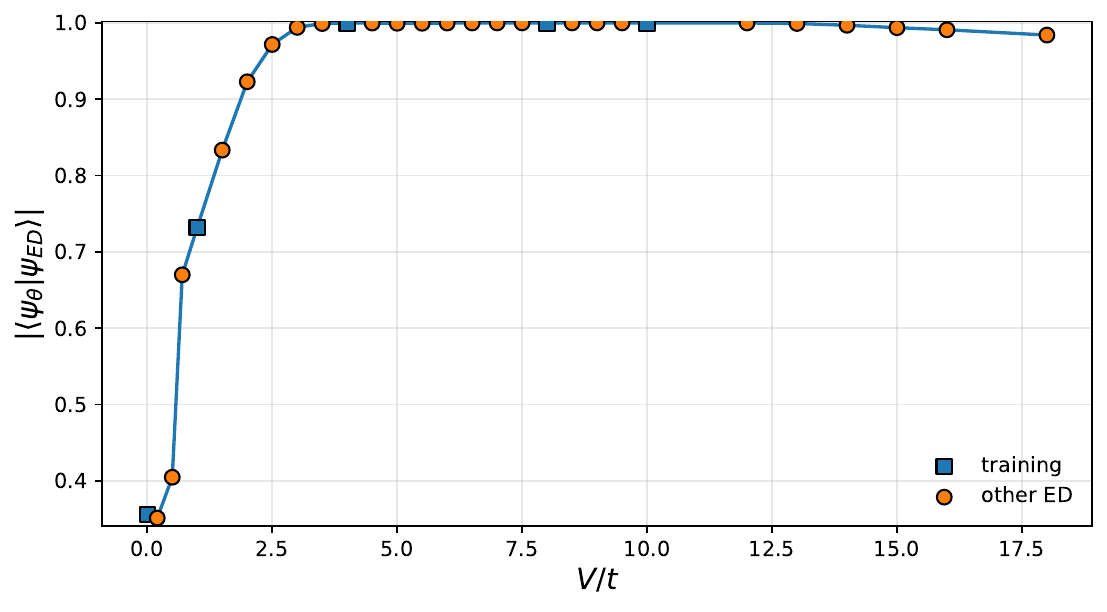}
\caption{\textbf{Overlap between the VMC optimized state and an ED
ground state representative.}
The upper panel shows the overlap magnitude for $N=6$, corresponding to
the energy comparison in Fig.~7. The lower panel
shows the overlap magnitude for the half filled $N=8$ system whose
energy is reported in the main text. Squares identify training
Hamiltonians and circles identify the remaining ED comparison points.
At $N=6$, the overlap varies strongly and can be small even where the
energy is extremely accurate. At $N=8$, the overlap approaches unity
throughout most of the interacting regime.}
\label{figVMCOverlapComparison}

\end{figure}

The upper panel of Fig.~\ref{figVMCOverlapComparison} reveals that the
excellent $N=6$ energy agreement in Fig.~\ref{figVMCEnergy} does not coincide with uniformly large
overlap. The overlap magnitude varies nonmonotonically with $V/t$, generally
remaining below approximately $0.7$ and approaching zero at several
parameters. These variations occur at both training and test Hamiltonians.
They therefore do not indicate a conventional failure of interpolation.
Instead, they expose the different information contained in energy
minimization and direct wave function supervision.

To make this distinction explicit, let
${\ket{\phi_m}}$ be the exact eigenstates of a given Hamiltonian with
energies $E_m$, ordered so that $E_0$ is the ground state energy. A
normalized variational state can be expanded as
\begin{equation}
\ket{\Psi_{\theta}}
=
\sum_m c_m\ket{\phi_m}.
\end{equation}
Its energy error is then
\begin{equation}
E_{\theta}-E_0
=
\sum_{m>0}
|c_m|^2
\left(E_m-E_0\right).
\label{eqEnergySpectralExpansion}
\end{equation}
Equation~\eqref{eqEnergySpectralExpansion} shows that the energy constrains
the distribution of the variational state across the spectrum, but does not
directly require the state to coincide with one selected eigenvector. A
state can have modest overlap with $\ket{\phi_0}$ and still have an energy
very close to $E_0$ when its remaining weight lies within a degenerate or
closely spaced low energy subspace.

This distinction is particularly important when the exact ground state is
degenerate or nearly degenerate. ED returns one representative from this
subspace, while energy minimization is free to converge to another linear
combination with the same or nearly the same energy. The overlap with the
chosen ED vector may then be small or even vanish although the VMC state
belongs to the physically relevant low energy manifold. The $N=6$ results
are consistent with this behavior. The energy varies smoothly and remains
accurate while the overlap with one exact representative changes sharply.

A more appropriate diagnostic in this situation is the total overlap with a
low energy subspace,
\begin{equation}
\mathcal{O}*{\mathrm{low}}(\epsilon)
=
\mel{\Psi*{\theta}}
{\hat P_{\epsilon}}
{\Psi_{\theta}},
\qquad
\hat P_{\epsilon}
=
\sum_{E_m-E_0\leq\epsilon}
\ket{\phi_m}\bra{\phi_m},
\label{eqLowEnergyOverlap}
\end{equation}
rather than the overlap with a single state. The energy result implies that
the VMC wave function has substantial weight in states with small
excitation energy, even when the single state overlap shown in
Fig.~\ref{figVMCOverlapComparison} is modest.

The lower panel of Fig.~\ref{figVMCOverlapComparison} presents a
complementary regime. At half filling, the overlap becomes nearly unity for
$V/t\gtrsim3$ and remains large throughout the strongly interacting region.
At weak coupling, however, the overlap is significantly smaller. This is
consistent with the greater difficulty of learning the detailed Fermi
surface structure near the noninteracting limit. The comparison between the
two particle sectors demonstrates that the relation between overlap and
energy depends on the spectral structure of each Hamiltonian. Neither
quantity should be interpreted as a universal substitute for the other.

For a unique ground state separated by a finite gap
$\Delta=E_1-E_0$, Eq.~\eqref{eqEnergySpectralExpansion} gives
\begin{equation}
E_{\theta}-E_0
\geq
\Delta
\left(1-|c_0|^2\right).
\label{eqEnergyGapBound}
\end{equation}
A sufficiently small energy error then enforces large fidelity. When the gap
is small, however, this bound becomes weak and many distinct wave functions
can possess nearly identical energies. Conversely, a large overlap does not
always guarantee a uniformly small absolute energy error when the many body
spectral width grows strongly with the interaction. A small residual weight
on high energy states can contribute appreciably to the energy even when
the total overlap remains close to one.

This separation has direct consequences for observable prediction. Energy
minimization constrains one expectation value and preferentially suppresses
wave function errors according to their excitation energies. It does not
uniformly constrain density correlations, structure factors, currents,
entanglement measures, or other operators. Two states with nearly identical
energies can give different values for these observables, particularly when
they occupy different symmetry sectors or different combinations within a
low energy manifold. Thus, the favorable scaling of VMC does not by itself
guarantee that all physical observables are learned with the same accuracy
as the energy.

Direct overlap training provides the complementary information. When exact
states are available, maximizing overlap constrains the complete amplitude
and phase structure of the wave function rather than a single expectation
value. The resulting model can predict arbitrary equal time observables
without introducing a separate objective for each one. Its limitation is
that producing complete target states requires ED or another solver and
therefore cannot be continued indefinitely with increasing system size. Nevertheless, there is a clear accuracy improvement that the overlap method provides versus VMC that builds a strong case to combine both in building a strong, broad foundation model for physical systems.

The combined results suggest a natural foundation model strategy. Exact
wave functions on smaller systems should be used to anchor the model through
overlap training, providing dense supervision of amplitudes, phases,
correlations, and symmetry structure. Variational energy minimization can
then extend the same shared representation to larger systems where complete
wave function labels are unavailable. The overlap objective supplies
physical information that energy alone may miss, while the VMC objective
removes the Hilbert space limitation of exact supervision
\cite{CarleoTroyer2017,Sorella2017,Rende2025}.

In practice, the overlap trained model can initialize VMC calculations at
larger sizes or both objectives can be optimized together over different
parts of the training ensemble. The excellent $N=6$ energy agreement shows
that the unsupervised component can remain accurate away from half filling.
The overlap comparisons show why exact wave function data should
nevertheless be retained wherever it is computationally available.
Together, these two objectives provide a scalable route toward foundation
models that preserve detailed wave function information on tractable
systems while extending variationally into regimes beyond ED.

\section{IV. ABLATIONS}
\label{app:ablations}

\begin{figure}
    \centering
    \includegraphics[width=\linewidth]{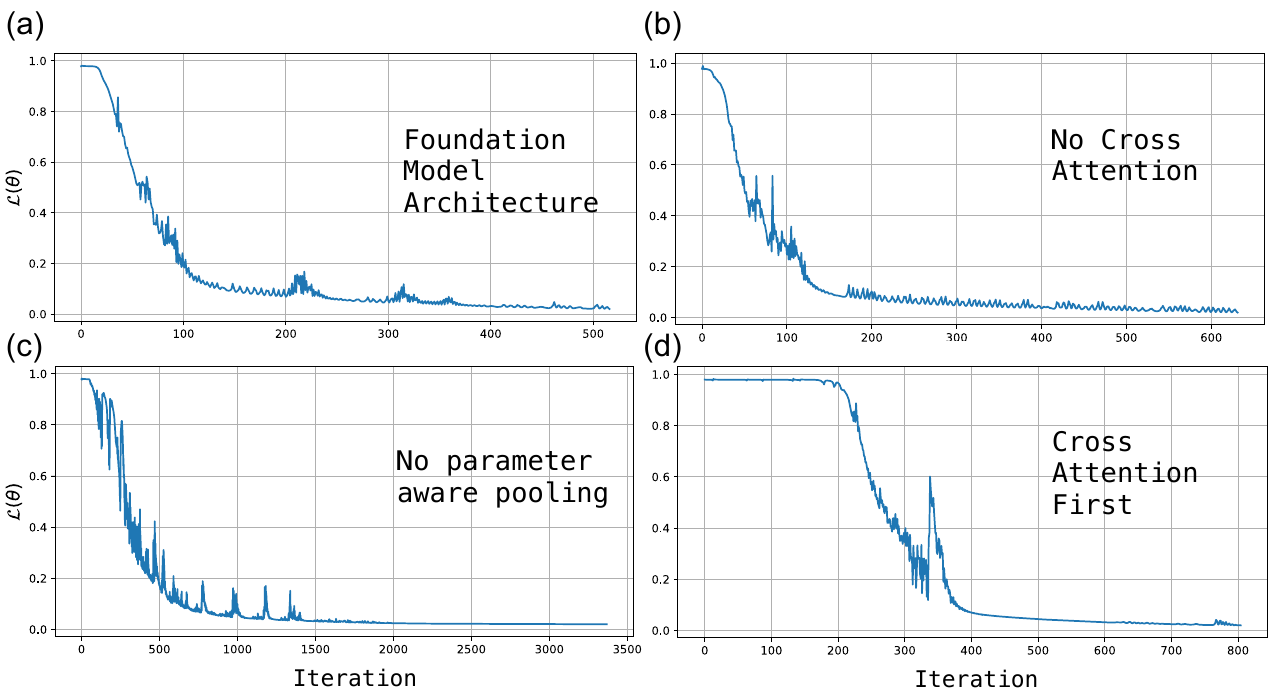}
    \caption{\textbf{Loss curves for 1d, 16 site systems at half filling.} {5 Hamiltonians are trained ($V/t$ = 0.0, 1.0, 4.0, 8.0, 16.0). Our foundation model architecture learns the best and most efficiently.}}
    \label{fig:ablations}
\end{figure}

To recap from the main text, our network takes a single basis token $\boldsymbol X = \{\boldsymbol x_i\}_{i=1}^{L}$ (binary or real features derived from a basis configuration) and a global parameter token $\boldsymbol \lambda$ that encodes the Hamiltonian couplings (e.g., $V/t$, filling). Tokens are first embedded, then processed by a self–attention (SA) block, a single cross–attention (CA) block in which $\boldsymbol \lambda$ attends to the site stream to tag each site with parameter information, followed by additional SA layers and a parameter–aware pooling (PAP) stage that compresses the site dimension conditioned on $\boldsymbol \lambda$ to the predicted complex amplitude of the basis token:
\[
\text{sites} \xrightarrow{\text{Embed}} 
\underbrace{\text{SA}}_{\text{context among sites}} \;\rightarrow\;
\underbrace{\text{CA}(\lambda \!\to\! \text{sites})}_{\text{one pass tagging}}
\;\rightarrow\; \text{SA}\;\; \xrightarrow{\text{PAP}(\cdot,\lambda)} \;\; \psi_\theta(\mathbf{n}_i,\lambda).
\]
The two design choices that mattered most empirically are (i) applying \emph{exactly one} CA pass, and (ii) using PAP instead of an unconditional pooling.

\subsection{A. Why cross–attention is used once after the first SA layer}
\label{app:ca-placement}

\smallskip
Cross attention is absolutely necessary as the network has a harder time learning the parameter to wavefunction map without it. Strictly speaking, the network still ``knows" about the parameters through Parameter-aware pooling (PAP), but the mapping becomes significantly stronger when the number of Hamiltonians increases. Since we are constrained to small number of Hamiltonians since we do not sample, this effect is not manifest in Fig. \ref{fig:ablations} ((a),(b)), however, the \~20\% improvement shown in Fig. \ref{fig:ablations} ((a),(b)) would be dramatically more visible as the number of Hamiltonians used to train increases. Thus the question becomes where to place the cross attention in the network.

Since the role of CA is to write a parameter label onto each site token, it just need to tags the basis token, so that subsequent SA layers can propagate and combine parameter–aware features. Repeating CA introduces redundant key–value copies of $\boldsymbol \lambda$ and competes with SA for capacity, increasing memory and gradient noise without adding information. On the other hand, placing CA before any SA forces the model to reconcile $\boldsymbol \lambda$ with unconditioned, local site embeddings; the result is poorly conditioned early features and slower representation alignment.

Fig. \ref{fig:ablations} ((b), (d)) shows that when CA moved to different points in the architecture, learning is stunted. Despite identical system and hyperparameters, training curves show:
(i) substantially higher training loss for the same wall–clock (longer plateaus), 
(ii) worse asymptotic fidelity and larger energy variance, and
(iii) degraded generalization across $V/t$ (non–monotone energy errors as in Fig.~\ref{fig:overlapvsEnergy}, but larger). 
Qualitatively, the network is presented with “too much conflicting information” too early, or not early enough: $\boldsymbol \lambda$ tries to modulate tokens that have not yet established spatial context through SA or have already been established to rigidly, leading to unstable attention patterns. 

Adding extra CA layers after every SA block yields negligible fidelity gains but higher memory usage, which forces a smaller network and worse learning overall. Again, since $\boldsymbol \lambda$ is global and low–entropy compared to the site stream, a single tag suffices; subsequent CA passes mostly re–encode the same label.

\subsection{B. Parameter–aware pooling (PAP)}
\label{app:pap}

Standard pooling discards information orthogonal to its fixed projection, potentially losing essential connections made during training. To ameleriote this issue, we compress in a way that is \emph{conditioned} on $\boldsymbol \lambda$. Concretely, given site features $\boldsymbol X\in\mathbb{R}^{L\times f}$ and a parameter embedding $p_\lambda\in\mathbb{R}^{f_\lambda}$,PAP computes a set of pooling weights and/or a pooling query via an MLP:
\begin{equation}
\mathrm{PAP}(\boldsymbol X,\boldsymbol \lambda) 
=\text{MLP}(\{\mathbf{X}, \boldsymbol \lambda\})
\label{eq:pap}
\end{equation}
followed by a linear projection over the feature dimension to produce the complex amplitude.

\smallskip
Replacing PAP by mean pooling or a linear projection (independent of $\lambda$) slows optimization and degrades the final loss. Fig. \ref{fig:ablations} (c) demonstrates the importance of this implementation. Without PAP, the network's ability to learned is severely diminished unilaterally. Most impressively, there are minimally added parameters used during this pooling strategy, making it highly effective while also highly efficient.

To summarize, while CA is necessary for parameter conditioning, it only needs to be injected at one point, which we report is most efficient and effective after the first attention layer. The importance of CA is highlighted in the calculation of a more complex Hamiltonian parameter space, like the one calculated for the order parameter heatmap (Fig. \ref{fig:heatmap}), as without CA, the network cannot even converge across the $\boldsymbol \lambda = (V/t, N)$ grid.

\section{V. RESOLUTION OF RESIDUAL DEGENERACIES}

In order to resolve the additional $D_4$ point group symmetries of the square lattice, we diagonalize the degenerate subspace described by the symmetry subspace operator \(\hat H_{\mathrm{sym}}\). Diagonalizing \(\hat H_{\mathrm{sym}}\) resolves the $C_4$ rotation and reflection degeneracy of the square lattice. Concretely, we construct the many-body fourfold rotation operator about the $z$-axis, $\hat{R}_{\pi/2}$, as well as the mirror operators $\hat{S}_x$ and $\hat{S}_d$, corresponding to reflections about the $\hat{x}$ axis and the diagonal $d$ axis, respectively. Explicitly,
\begin{equation}
        \hat H_{\mathrm{sym}}
    \;=\;
    \lambda_{sym}\!\left(2\mathbbm{1} - r_\star(\hat R_{\pi/2}+\hat R_{\pi/2}^{\dagger})\right)
    \;+\; \mu\!\left((\mathbbm{1}-s_x\,\hat S_x) + (\mathbbm{1}-s_d\,\hat S_d)\right),
\end{equation}
with targets \((r_\star,s_x,s_d)\) fixed by the desired one or two-dimensional irrep \(\Gamma\in\{A_1,A_2,B_1,B_2,E\}\). 
We target a given irrep to select which degenerate sector to choose our ground state from. 
This guarantees that the wavefunction is from the same degenerate sector across the parameter manifold, ensuring the smooth map, $\boldsymbol \lambda \rightarrow \Psi_\theta(\boldsymbol \lambda)$, required for foundation model generalization. In practice we set $\lambda_{sym}, \mu = 10^{-2}$, which allows us to resolve the degeneracies for all sectors across the $(V/t,N)$ grid.

\section{VI. EXTENDED DATA}

In this section we include additional figures that provide more insight and visibility into our results.

\subsection{A. Energy landscape over the full $(V/t,N)$ grid}
Extended Data Fig.~\ref{fig:ext-energy} shows the total variational energy $E_\theta/t$ across the full $(V/t,N)$ grid. The relative error $|E_\theta-E_{ED}|/|E_{ED}|$ is within about $5\%$ over the domain. As the interaction scale increases, the absolute deviation plateaus while $|E_{ED}|$ decreases, which enlarges the relative error; normalized errors remain smooth across the grid, consistent with the fidelity–energy width discussion in Sec.~\ref{sec:energy_from_overlap}.
\begin{figure}[t]
  \centering
  \includegraphics[width=\linewidth]{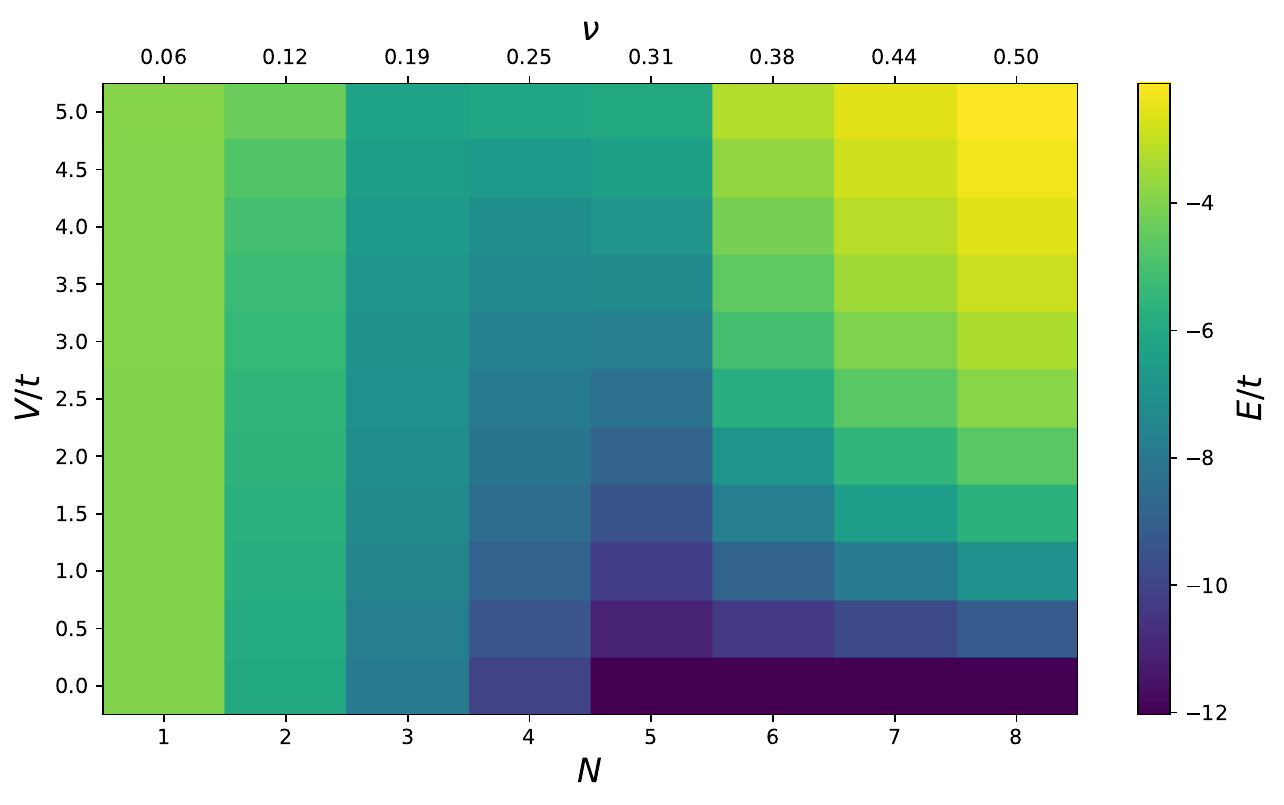}
  \caption{\textbf{Variational energy $E_\theta/t$ across $(V/t,N)$}. Here, energy is computed by the Rayleigh quotient of the trained state, and relative error to ED is within approximately $5\%$.}
  \label{fig:ext-energy}
\end{figure}

\subsection{B. Pair binding energy across $(V/t,N)$}
We report the pair binding energy (PBE) doping the ground state at filling $N$ on the nearly half-filled grid $N\in\{2,\dots,7\}$ for a $4\times4$ lattice and $V/t\in[0,5]$,
\begin{equation}
E_b(N;V/t)=E_{\theta}(N;V/t)+E_{\theta}(N+2;V/t)-2\,E_\theta(N+1;V/t) ,
\label{eq:pbe-def-app}
\end{equation}

\begin{figure}[t]
  \centering
  \includegraphics[width=\linewidth]{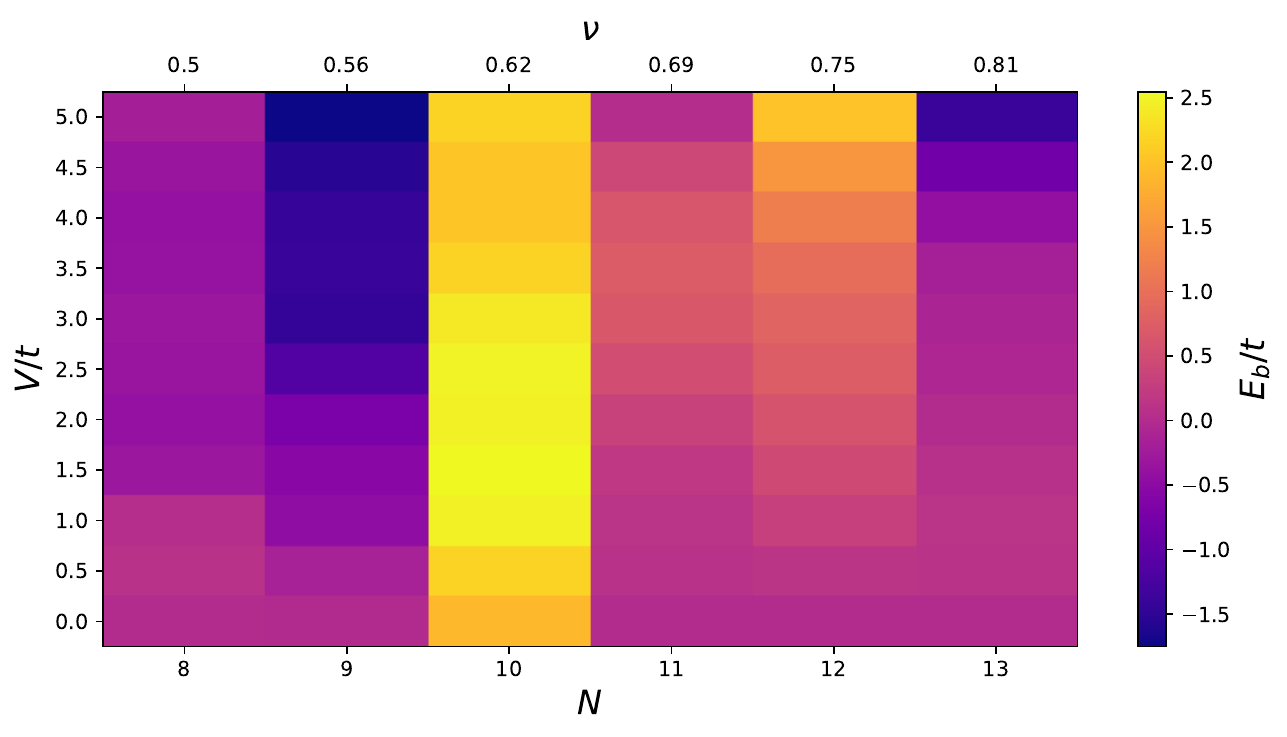}
  \caption{\textbf{Pair binding energy $E_b/t$ on $N\in\{2,\dots,7\}$ and $V/t\in[0,5]$ for a $4\times4$ lattice.}}
  \label{fig:ext-pbe}
\end{figure}

where $E_{\theta}(N;V/t)$ is the ground-state energy produced by the network in the fixed-$N$ sector and the same Hamiltonian and symmetry constraints are used for $N$, $N+1$, and $N+2$. 
Negative $E_b$ indicates an effective attraction. 
The heat map in Extended Data Fig. S7 shows the expected crossover from kinetic-dominated to interaction-dominated regimes as $V/t$ increases; the magnitude of $E_b$ grows near half filling while remaining small away from it.


\subsection{C. Sanity checks}
For every panel we verify: normalization of the variational state; particle-number sum rule from the one-body density matrix, $\sum_i\langle n_i\rangle=N$; two-particle trace $\sum_{i,j}\langle n_i n_j\rangle=N(N-1)$ using diagonal two-body elements; and agreement between ED eigenpairs and direct expectations $\langle\psi_0|H|\psi_0\rangle$. All checks pass within machine precision for each system.

\section{VII. Fock basis states}

\begin{figure}
    \centering
    \includegraphics[width=0.4\linewidth]{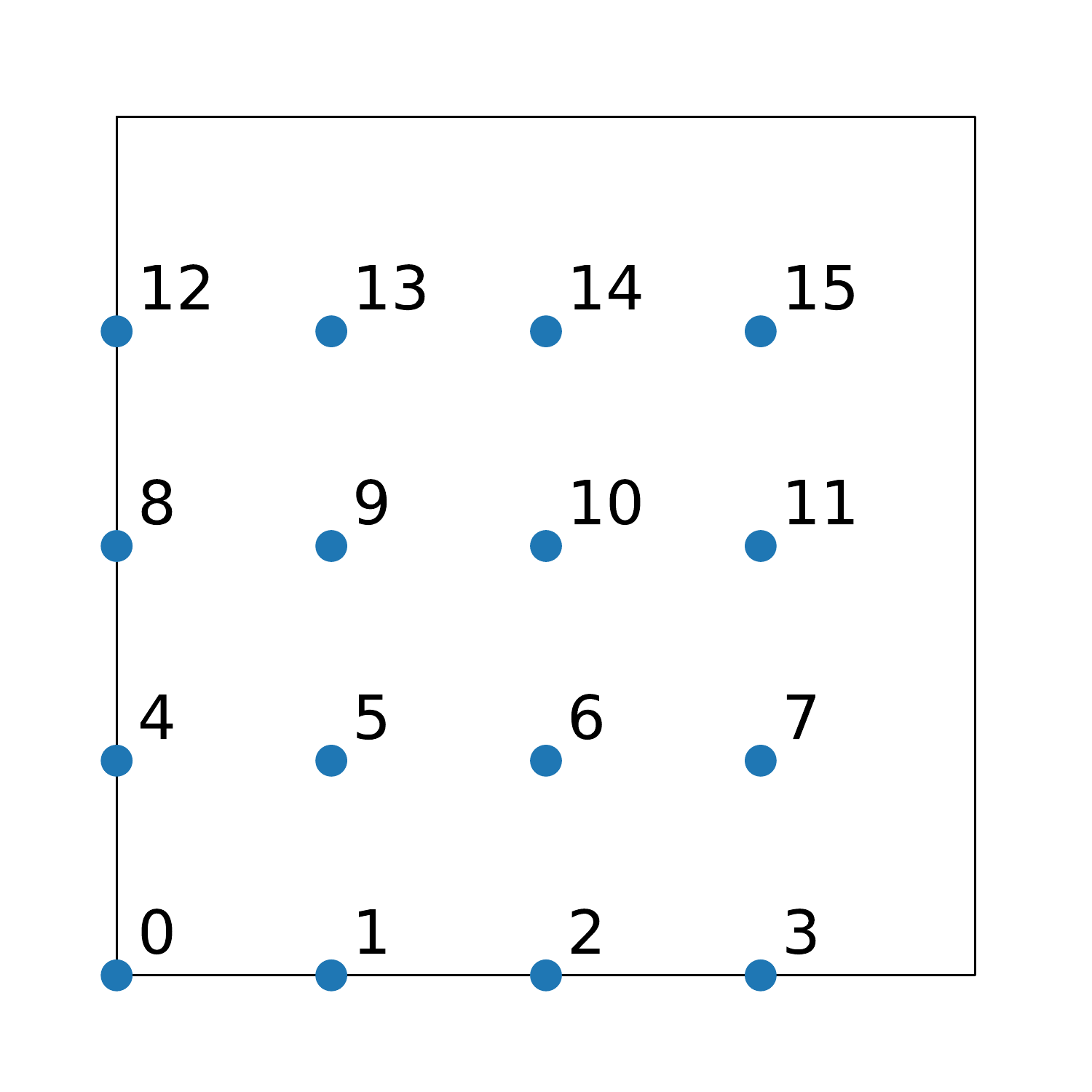}
    \caption{Lattice-site ordering adopted for defining Fock basis states on the square lattice.}
    \label{fig:squatrlattice}
\end{figure}

The Fock state is defined as follows: 
\begin{equation}\label{app:fockstate}
\ket{\mathbf n}  = \left(c^\dagger_1\right)^{n_1}\cdots \left(c^\dagger_L\right)^{n_L}\ket{0},
\end{equation}
where $\ket{0}$ is the state with no particles, $n_i=0,1$ is the fermion occupation number and $c^\dagger_i$ is the creation operator which adds a fermion in state $i$. 
The definition in Eq.~\eqref{app:fockstate} implicitly assumes a specific ordering of the lattice sites. In our calculations on the two-dimensional square lattice, we adopt the site ordering shown in Fig. S8.

\end{widetext}

\end{document}